\documentclass[hyper,11pt,paper]{article}
\usepackage{jheppub}
\usepackage{graphicx,amsmath,amssymb,multirow,array,bm,mathrsfs}
\usepackage{epsf,amsfonts}
\usepackage[numbers,sort&compress]{natbib}
\usepackage{hyperref}
\newcommand\beq{\begin{equation}}
\newcommand\eeq{\end{equation}}
\newcommand\be{\begin{equation}}
\newcommand\ee{\end{equation}}

\title{Revisiting non-relativistic limits}

\preprint{YITP-SB-14-50}

\author[a]{Kristan Jensen,}
\author[b]{Andreas Karch}

\affiliation[a]{C.N. Yang Institute for Theoretical Physics, SUNY Stony Brook, Stony Brook, NY 11794-3840, USA}
\affiliation[b]{Department of Physics, University of Washington, Seattle, WA 98195, USA}
\emailAdd{kristanj@insti.physics.sunysb.edu}
\emailAdd{akarch@uw.edu}

\abstract{We show that the full spurionic symmetry of Galilean-invariant field theories can be deduced when those theories are the limits of relativistic parents. Under the limit, the non-relativistic daughter couples to Newton-Cartan geometry together with all of the symmetries advocated in previous work, including the recently revived Milne boosts. Our limit is a covariant version of the usual one, where we start with a gapped relativistic theory with a conserved charge, turn on a chemical potential equal to the rest mass of the lightest charged state, and then zoom in to the low energy sector. This procedure gives a simple physical interpretation for the Milne boosts. Our methods even apply when there is a magnetic moment, which is known to modify the non-relativistic symmetry transformations. We focus on two examples, taking the non-relativistic limits of scalar field theory and hydrodynamics.}

\begin{document}
\maketitle

%%%%%%%%%%%%%%%%%%%%%%%%%%%%%%%%%%%%%%%%%

%%%%%%%%%%%%%%%%%%%%%%%%%%%%%%%%%%%%%%%%%
\section{Introduction}
%%%%%%%%%%%%%%%%%%%%%%%%%%%%%%%%%%%%%%%%%

Starting with the pioneering work of Son and Wingate~\cite{Son:2005rv}, diffeomorphism invariance has emerged as a powerful new tool to constrain the low-energy dynamics in Galilean invariant non-relativistic (NR) quantum systems even in the presence of strong interactions. This class of theories includes systems of obvious importance such as strongly correlated electrons. Already in~\cite{Son:2005rv}, these considerations were used to derive new constraints on transport in the unitary Fermi gas, but since then the same approach has been successfully applied to systems as diverse as Hall states~\cite{Hoyos:2011ez,Andreev:2013qsa,Son:2013rqa,Geracie:2014nka} or chiral \cite{Hoyos:2013eha,Moroz:2014ska} and standard \cite{Brauner:2014jaa} superfluids.

In a quantum field theory, we should regard the background metric, gauge fields, and other sources as coupling constants. Then coordinate reparameterizations (which are usually referred to as diffeomorphisms) are a spurionic symmetry transformation, under which the couplings change but the action is left invariant. See e.g.~\cite{Janiszewski:2012nb} for a discussion. From this one can derive powerful constraints on how these coupling constants appear in the low energy effective action. This is particularly useful in gapped phases, where the low energy effective action solely depends on the background parameters. Of course, most of the low-derivative terms in said effective action are unphysical, as they are indistinguishable from local counterterms. In this regard, Chern-Simons terms and other topological terms are special, as local counterterms can only shift their couplings in a discrete way. In this setting, the authors of~\cite{Hoyos:2011ez} showed that for the Hall effect, the standard electromagnetic Chern-Simons term is not invariant under the ``non-relativistic diffeomorphisms'' of~\cite{Son:2005rv}, and so must be accompanied by other terms in the action that in principle give rise to measurable effects.

In a relativistic quantum field theory, these background coupling constants are just the spacetime metric and, when the theory has a conserved charge, a background gauge field. We can think of the background metric and gauge field as sources for energy-momentum and global symmetry currents. The analogous structure for a Galilean-invariant non-relativistic theory has only been understood recently. The authors of~\cite{Son:2013rqa,Geracie:2014nka} have argued that, for a Galilean-invariant theory, the sources for energy, momentum \&c, comprise a version of  ``Newton-Cartan'' (NC) geometry, which we review in more detail below. That is, Galilean theories ought to couple to a NC structure, rather than a usual metric. The fields in a NC structure are all tensors, and so it is easy to couple these sources to a Galilean theory in a way that manifests the coordinate reparameterization and gauge invariances. The full symmetries however include one more crucial new ingredient~\cite{Jensen:2014aia}, the so-called Milne boosts (so-called in segments of the NC literature, e.g.~\cite{Duval:1983pb}). The Milne boosts ensure that there is some redundancy in the NC data, and its Ward identity equates momentum and particle number currents. Fixing the Milne symmetry leads to the ``diffeomorphism invariance'' of Son and Wingate~\cite{Son:2005rv} and~\cite{Son:2013rqa,Geracie:2014nka}. Correspondingly, the ``non-relativistic diffeomorphisms'' appearing in those papers acted in a somewhat unusual fashion which explicitly depended on a choice of coordinates, as they always had to be accompanied by a compensating Milne boost to retain the Milne-fixing condition.\footnote{Some alternative proposals for coupling non-relativistic field theories to Newton-Cartan geometry may be found in~\cite{Banerjee:2014pya,Banerjee:2014nja,Hartong:2014pma}. Their relation to the formulation in~\cite{Son:2005rv,Geracie:2014nka,Jensen:2014aia} is not entirely clear at this time.}
 
In this note, we show how NC geometry and the Milne boosts naturally arise from a non-relativistic limit. We start with a gapped relativistic parent with a conserved charge, and couple it to an ordinary spacetime metric as well as a background gauge field. We then take a NR limit by turning on a chemical potential equal to the rest energy of the lightest charged relativistic particles, followed by a $c\to\infty$ limit, where $c$ is the speed of light. We find that the NC data naturally arises from the relativistic background. Even the connections for NC geometry, the analogues of the Levi-Civita connection built from an ordinary metric, emerge in the $c\to\infty$ limit. The Milne boost invariance also arises naturally. In order to perform our non-relativistic limit, we split our relativistic metric and gauge field into a leading and a subleading term. For example, in the relativistic background gauge field the leading piece is the $\mathcal{O}(c^2)$ chemical potential $mc^2$. The $\mathcal{O}(c^0)$ part of the relativistic gauge field becomes the background gauge field that couples to the particle number current of the NR daughter theory. This split, however, is not unique. We can always redefine the coefficient of the $\mathcal{O}(c^2)$ term by a small $\mathcal{O}(c^{-2})$ correction, and then redefine the order $\mathcal{O}(c^0)$ NR gauge field by an equal opposite amount, which just leaves the relativistic gauge field unchanged. This reshuffling freedom between leading and subleading terms gives rise to the Milne boost invariance.

We focus on two examples for which we perform this limit: free scalar field theory, and relativistic hydrodynamics. In each case our limit is very similar in spirit to the standard large $c$ limits. NC geometry and Milne boosts emerge in the NR limit of both examples, and we expect that this is true more generally.

Part of our motivation is to better understand non-relativistic `t Hooft anomalies. In relativistic theories, anomalies provide a window into non-perturbative physics; they must be matched across scales and so place strong constraints on renormalization group flow. Anomalies are also robust against strong correlations and even disorder. The understanding of anomalies in non-relativistic theories, certainly relevant for the edges of topologically non-trivial phases, is still emerging. To properly classify those anomalies, as well as to deduce whether they are robust, one must first understand the potentially anomalous symmetries.

This note is organized as follows. In the next Section we review the essentials of NC geometry and Milne boosts, and then show how to obtain them from the NR limit of free scalar field theory. In Section~\ref{S:magnetic} we then study free scalars in $d=2+1$ space-time dimensions, where a magnetic moment is allowed by both the relativistic and non-relativistic symmetries. Such terms have played a prominent role in recent work on NC geometry~\cite{Son:2013rqa,Geracie:2014nka,Jensen:2014aia}, since they facilitate a massless limit of a Hall system, whereby the lowest Landau level decouples from the others. For us the main role they play is that they give an interesting testing ground for our construction. In the NR theory, the magnetic moment necessitates a modification of the Milne symmetry. We derive these modified transformation rules, for a particular form of the magnetic moment, from a relativistic parent following the same procedure as in Section~\ref{S:RtoNR}. This exercise also helps us to shed some light on the limitations of our approach: not every NR action consistent with the symmetries of the NC geometry can be realized by a relativistic parent following our prescription. In Section~\ref{S:hydro} we study the NR limit of relativistic hydrodynamics in some detail. We show how to systematically obtain the NR constitutive relations, Ward identities, and entropy current from those of the relativistic parent, and in so doing our construction naturally matches the recent covariant presentation of NR hydrodynamics in~\cite{Jensen:2014ama}. We conclude with some questions for the future in Section~\ref{S:conclude}.

%%%%%%%%%%%%%%%%%%%%%%%%%%%%%%%%%%%%%%%%%
\section{The non-relativistic limiting procedure}
\label{S:RtoNR}
%%%%%%%%%%%%%%%%%%%%%%%%%%%%%%%%%%%%%%%%%

%%%%%%%%%%%%%%%%%%%%%%%%%%%%%%%%%%%%%%%%%
\subsection{Newton-Cartan from a relativistic parent}
%%%%%%%%%%%%%%%%%%%%%%%%%%%%%%%%%%%%%%%%%

One way to describe Newton-Cartan (NC) geometry in $d$ spacetime dimensions is in terms of three pieces of data~\cite{Jensen:2014aia} (see also~\cite{Duval:1984cj,Christensen:2013rfa})
\beq
A_{\mu}\,, \quad n_{\mu}\,, \quad h_{\mu \nu}\,.
\label{nc}
\eeq
Here, $A_{\mu}$ is a $U(1)$ gauge field which couples to particle number and $h_{\mu\nu}$ is a symmetric, positive semi-definite rank $d-1$ tensor. $n_{\mu}$ and $h_{\mu\nu}$ are almost arbitrary: we require that
\beq
\label{E:defineGamma}
\gamma_{\mu\nu}\equiv n_{\mu}n_{\nu}+h_{\mu\nu}\,,
\eeq
is a positive-definite, rank$-d$ tensor. The NC data transform under diffeomorphisms and gauge transformations in the standard way. In addition, $A_{\mu}$ and $h_{\mu\nu}$ shift under Milne boosts as we explain below.

Since $h_{\mu \nu}$ has a single zero eigenvalue, we can introduce the unique corresponding eigenvector $v^{\mu}$ such that
\beq
\label{E:defineV}
h_{\mu \nu} v^{\nu} =0\,, \quad n_{\mu} v^{\mu} =1\, .
 \eeq
Roughly speaking, $n_{\mu}$ and $v^{\mu}$ define the time direction and $h_{\mu \nu}$ is the spatial metric. $v^{\mu}$ allows us to formulate two auxiliary quantities
\beq
 \label{others} P^{\mu}_{\nu} = \delta^{\mu}_{\nu} - v^{\mu} n_{\nu}\,, \quad h^{\mu \rho} h_{\rho \nu} = P^{\mu}_{\nu}\,. \eeq
Note that $h^{\mu \nu}$ is not the inverse of $h_{\mu \nu}$. $h^{\mu \nu}$ is defined by the second equation in \eqref{others}.

Let us see how this data naturally arises in the NR limits of relativistic theories. The relativistic theory couples to a background metric $g_{\mu \nu}$ as well as of a background gauge field $C_{\mu}$. As in \cite{Janiszewski:2012nb} we will start with a free complex scalar,
\beq
\label{E:Rfree}
S = - \int d^{d-1}x dt \,  \sqrt{-g} \, \left (\frac{1}{2}g^{\mu \nu} {\cal D}_{\mu} \Phi^* {\cal D}_{\nu} \Phi+ \frac{c^2 m^2}{2} |\Phi|^2 \right )\,.
\eeq
The covariant derivative is defined as usual as ${\cal D}_{\mu} \Phi = \partial_{\mu} \Phi - i C_{\mu} \Phi$. If we start in flat Minkowski space,
\beq
g_{\mu \nu} dx^{\mu} dx^{\nu} = - c^2 dt^2 + d \vec{x}^2 \,,
\eeq
the procedure to get a NR limit can be accomplished by three simple steps.
\begin{enumerate}
\item Turn on a background chemical potential via $C_{\mu} = m c^2 \delta_{\mu}^t + mA_{\mu}$, where $A_{\mu}$ is the NR gauge field.
\item Rescale the field as $\phi = \sqrt{mc} \, \Phi$ to remove all overall factors of $c$.
\item Send $c \rightarrow \infty$.
\end{enumerate}
We implicitly study field configurations where $\Phi$ and $A_{\mu}$ vary of length scales which do not scale with $c$. In this limit the $|\partial_t\Phi|^2$ term drops out as $g^{tt} = -1/c^2$. The $g^{tt} C_t C_t$ term is $\mathcal{O}(c^2)$, but cancels against the rest mass: indeed, we chose the chemical potential to compensate the rest energy. Particles have a kinetic energy $p^2/(2m)$, anti particles have energy $2 m c^2 + \ldots$ and completely decouple. The theory reduces to the standard NR action
\beq
\lim_{c\to\infty}S = \int d^{d-1} x dt \left [ \frac{i}{2} (\phi^*D_t \phi - \phi D_t \phi^*)  - \frac{\delta^{ij}}{2m} D_i \phi^* D_j \phi \right ]\,.
\eeq
The NR covariant derivative $D_{\mu}$ involves the NR gauge field $A_{\mu}$, $D_{\mu} \phi = \partial_{\mu}\phi - i  mA_{\mu} \phi$. This procedure of getting NR theories from relativistic ones by canceling the rest mass via a chemical potential, and then sending $c\to\infty$, has also recently been implemented successfully in hydrodynamics~\cite{Kaminski:2013gca}.

We now show that if we instead start with the most general relativistic background that allows for a NR limit, we obtain exactly the NC structure as defined in \eqref{nc}. In the limit above, it was important that the $dt^2$ piece in the metric came with an extra prefactor of $c^2$, so that in the inverse metric the corresponding $1/c^2$ term killed the two-time derivative term in the action~\eqref{E:Rfree}. If we introduce a covariant vector $n_{\mu}$ in order to pick the time direction, we write the relativistic metric as\footnote{A somewhat different approach to obtain NR theories as a limit of relativistic parents has recently been put forward in \cite{Andreev:2014gia}.}
\beq
\label{metric} g_{\mu \nu} = - c^2 n_{\mu} n_{\nu} + h_{\mu \nu} \,,
\eeq
where $h_{\mu \nu}$ has rank $d-1$ so we do not overcount. Correspondingly, we can introduce all the quantities $v^{\mu}$, $h^{\mu \nu}$ and $P^{\mu}_{\nu}$ as above. The inverse metric is
\beq
g^{\mu \nu} =  - \frac{1}{c^2} v^{\mu} v^{\nu}+h^{\mu \nu}\,.
 \eeq
This already has the correct feature that the only two derivative terms in~\eqref{E:Rfree} that survive the $c \rightarrow \infty$ limit will come with $h^{\mu \nu}$, as the terms with $v^{\mu}$ are suppressed by an inverse power of $c^2$. In order to cancel the rest mass, we need to introduce a background gauge field, but now we allow the $mc^2$ term to be along a covector $b_{\mu}$
\beq
\label{E:Rgauge}
C_{\mu} = m c^2 b_{\mu} + mA_{\mu}\,.
\eeq
Demanding that $S$ is regular under our $c\to\infty$ limit, we find that
\beq h^{\mu \nu} b_{\nu}=0\,, \quad b_{\mu} v^{\mu} = \pm 1\, ,
\eeq
and taking the $+$ convention, by~\eqref{E:defineV} and~\eqref{others} this fixes
\beq
\label{nbrel} b_{\mu} = n_{\mu}\, .
\eeq

Plugging in the forms~\eqref{metric} and~\eqref{E:Rgauge} for $g_{\mu \nu}$ and $C_{\mu}$ back into the action, rescaling $\Phi$ to $\phi$ and taking the $c \rightarrow \infty$ limit we arrive at
\beq
\label{freeaction}
\lim_{c\to\infty} S = \int d^{d-1} x dt \sqrt{\gamma} \left [ \frac{iv^{\mu}}{2} (\phi^* D_{\mu} \phi - \phi D_{\mu} \phi^*)  - \frac{h^{\mu \nu}}{2m} D_{\mu} \phi^* D_{\nu} \phi \right ].
\eeq
where $\sqrt{\gamma} = \sqrt{-g}/c = \sqrt{\det(\gamma_{\mu\nu})}$ and $\gamma_{\mu\nu}$ is defined in~\eqref{E:defineGamma}. This is exactly the right action of a NR scalar field on an arbitrary NC geometry, as argued in \cite{Jensen:2014aia}. It was shown there that $\sqrt{\gamma}$ is the correct volume element for a general NC geometry.

%%%%%%%%%%%%%%%%%%%%%%%%%%%%%%%%%%%%%%%%%
\subsection{Transformation properties and Milne boosts}
%%%%%%%%%%%%%%%%%%%%%%%%%%%%%%%%%%%%%%%%%

Since $n_{\mu}$, $h_{\mu \nu}$ and $A_{\mu}$ were all defined as ordinary covariant tensors, they transform in the usual way under coordinate reparameterizations. In addition, $A_{\mu}$ shifts under gauge transformations. So all the NC data transforms exactly as it should. In order to complete the comparison to \cite{Jensen:2014aia}, all we need to do is to obtain the Milne boosts.

What is the physical origin of the Milne boost invariance? In our NR limit this is very clear. We needed to split the relativistic fields, $g_{\mu \nu}$ and $C_{\mu}$, into a leading $\mathcal{O}(c^2)$ part determined by $n_{\mu}$, and subleading $\mathcal{O}(1)$ pieces $h_{\mu \nu}$ and $A_{\mu}$. Clearly this split is ambiguous. We can always redefine $n_{\mu}$ by an $\mathcal{O}(c^{-2})$ term, which can be compensated by an equal opposite shift in $h_{\mu\nu}$ and $A_{\mu}$ so as to leave $g_{\mu\nu}$ and $C_{\mu}$ unchanged. With this insight, we can identify the Milne shift as
\beq
\label{milne}
n_{\mu}\to n_{\mu}  - \frac{\Psi_{\mu}}{c^2}\,, \quad  A_{\mu} \to A_{\mu} + \Psi_{\mu}, \quad
h_{\mu \nu}\to h_{\mu\nu}  -(n_{\mu} \Psi_{\nu} + n_{\nu} \Psi_{\mu}) + \frac{1}{c^2} \Psi_{\mu} \Psi_{\nu}\,,
\eeq
which in the $c \rightarrow \infty$ limit reduces to
\beq
\label{milnef}
A_{\mu}\to A_{\mu} + \Psi_{\mu}, \quad
 h_{\mu \nu} \to h_{\mu\nu} - (n_{\mu} \Psi_{\mu} + n_{\nu} \Psi_{\mu})\,.
\eeq
In order to ensure that $h_{\mu \nu}$ remains rank $d-1$, we require
\beq
\label{Psi}
\Psi_{\mu} =  \psi_{\mu} - \frac{1}{2} n_{\mu} \psi^2\,,
\eeq
where $\psi_{\mu}$ is spatial, $v^{\mu}\psi_{\mu}=0$, and
\beq
 \psi^2 = h^{\mu \nu} \psi_{\mu} \psi_{\nu}\,.
\eeq
The new $h_{\mu \nu}$ still has a zero eigenvalue, and the new corresponding eigenvector is
\beq
 v^{\mu}\to v^{\mu} + h^{\mu \nu} \psi_{\nu}\,.
\eeq
These transformation rules exactly reproduce the ones derived for Milne boosts in~\cite{Jensen:2014aia}. This shows that the full NR action, with all background fields in the NC formalism, as well as their transformation laws, can be nicely understood from a relativistic parent theory in the $c \rightarrow \infty$ limit. Any NR action that arises as a NR limit along these lines from a relativistic parent is automatically invariant under Milne boosts, since the latter were engineered to leave the relativistic sources invariant. They simply amounted to a shift of some of the leading $\mathcal{O}(c^2)$ pieces into the subleading $\mathcal{O}(1)$ pieces in the NR limit.

%%%%%%%%%%%%%%%%%%%%%%%%%%%%%%%%%%%%%%%%%
\subsection{The connection}
%%%%%%%%%%%%%%%%%%%%%%%%%%%%%%%%%%%%%%%%%

The connection is an important ingredient in NC geometry. In our Lagrangian~\eqref{freeaction} all fields were scalars and so we did not have to commit to a connection. But in order to define covariant derivatives of general NC tensors, or to efficiently express the partition function $\mathcal{Z}$, one needs to also define a connection. In Riemannian geometry, there is a unique connection which can be defined just using the metric. In NC geometry, there are many connections that can be built from $(n_{\mu},h_{\mu\nu},A_{\mu})$. According to~\cite{Geracie:2014nka,Jensen:2014aia} there is a natural choice whereby one requires that the NC data $n_{\mu}$ and $h^{\mu \nu}$ are covariantly constant and the spatial torsion vanishes. The corresponding connection reads
\beq
\label{giconnection}
\Gamma^{\mu}{}_{\nu \rho} = v^{\mu} \partial_{\rho} n_{\nu} + (\Gamma_{h})^{\mu}{}_{\nu \rho} + h^{\mu \sigma}
n_{(\nu} G_{\sigma) \rho}\,.
\eeq
Here $(\ldots)$ denotes symmetrization with weight $1/2$, $G_{\mu\nu}$ is an arbitrary two-form, and
\beq
(\Gamma_h)^{\mu}{}_{\nu \rho} = \frac{1}{2} h^{\mu \sigma} (\partial_{\nu} h_{\rho \sigma} + \partial_{\rho} h_{\nu \sigma} -
\partial_{\sigma} h_{\nu \rho} )\,,
\eeq
can be thought of as the analog of the standard Christoffel symbols on $h_{\mu \nu}$. If we only use the NC data to define the connection, there are two natural choices for $G_{\mu\nu}$: $0$, which was taken in~\cite{Geracie:2014nka}, or $F_{\mu\nu}$, the field strength of $A_{\mu}$, which was taken in~\cite{Jensen:2014aia}. We follow~\cite{Jensen:2014aia} and take $G_{\mu\nu} = F_{\mu\nu}$.

Note that the first term in the connection is not symmetric under exchange of $\rho$ and $\nu$. So the price one had to pay for constancy of the NC data is the temporal torsion
\beq
T^{\mu}{}_{\nu \rho} = \Gamma^{\mu}{}_{\nu\rho} - \Gamma^{\mu}{}_{\rho\nu} = v^{\mu} F^n_{\rho \nu}\,, \qquad F^n_{\mu \nu} = \partial_{\mu} n_{\nu} - \partial_{\nu} n_{\mu}\,.
\eeq
We regard $F^n_{\mu\nu}$ as the analogue of a field strength for $n_{\mu}$. By construction, this connection is gauge-invariant. It is however not Milne-invariant. As shown in~\cite{Jensen:2014aia}, it is possible to define a manifestly Milne-invariant connection at the price of giving up gauge invariance. This Milne but not gauge-invariant connection $\Gamma_A$ is
\beq
\label{miconnection}
(\Gamma_A)^{\mu}{}_{\nu \rho} \equiv \Gamma^{\mu}{}_{\nu \rho} + h^{\mu \sigma}
( - A_{\sigma} \partial_{[\rho} n_{\nu]} + A_{\nu} \partial_{[\rho} n_{\sigma]} + A_{\rho} \partial_{[\nu} n_{\sigma]} )\,,
\eeq
where $[\ldots]$ denotes anti-symmetrization with weight $1/2$. Of course in the end we are interested in theories which are both gauge and Milne invariant, but at the level of the connection it is only possible to manifest one or the other. Last but not least, we can shift the connection by a term proportional to $v^{\mu} F^n_{\nu \rho}$ to obtain the torsionless version of the gauge but not Milne-invariant NC connection, and a similar shift can be applied to the Milne but not gauge invariant connection to obtain its torsionless version. It is this latter, torsionless Milne invariant connection that naturally arises when embedding the NC data into a lightlike reduction of a $d+1$-dimensional metric \cite{Jensen:2014aia} (see also~\cite{Christensen:2013rfa}).

We would like to see how this set of closely related connections arises from the NR limit. Our starting point is the standard Levi-Civita connection associated to the metric \eqref{metric}. We refer to this relativistic connection as $\Gamma_R$. Since $\Gamma_R$ is torsionless, we naturally obtain a torsionless connection from its NR limit. We presently demonstrate that we naturally obtain both the gauge-invariant and Milne-invariant torsionless connections from the $c\to\infty$ limit.

Organizing the terms in $\Gamma_R$ order by order in $c^2$, we find that the $\mathcal{O}(c^2)$ term is
\beq
(\Gamma_R)^{\mu}_{\nu \rho} = -c^2 h^{\mu \sigma}  n_{(\rho} F^n_{\nu) \sigma}  + {\cal O}(1)\,.
\eeq
To take the $c\to\infty$ limit, we need to make a tensorial redefinition of $\Gamma_R$ that eliminates this $\mathcal{O}(c^2)$ piece. There are at least two ways we can accomplish this. One is to add a tensor involving the relativistic vector potential $C_{\mu}$ and its field strength
\beq
\mathcal{F}_{\mu\nu} = \partial_{\mu}C_{\nu}-\partial_{\nu}C_{\mu} = m c^2 F^n_{\mu\nu} + mF_{\mu\nu}\,,
\eeq
which gives a manifestly Milne invariant connection, as all relativistic sources are manifestly invariant under Milne boosts. Of course the price to pay is that the resulting NR connection is not gauge-invariant. Instead, we can add a tensor involving $n_{\mu}$ and $\mathcal{F}_{\mu\nu}$. In this case the tensorial redefinition is not Milne-invariant, but instead is gauge-invariant. That is, we define
\beq (\Gamma^{(1)}_R)^{\mu}{}_{\nu \rho} = (\Gamma_R)^{\mu}{}_{\nu \rho} +\frac{1}{m} g^{\mu \sigma} n_{(\rho} \mathcal{F}_{\nu) \sigma}\,,
\quad \quad
(\Gamma^{(2)}_{R})^{\mu}{}_{\nu \rho} = (\Gamma_R)^{\mu}{}_{\nu \rho} + \frac{1}{m^2c^2}g^{\mu \sigma} C_{(\rho} {\cal F}_{\nu) \sigma}\,.
\eeq
Both $\Gamma^{(1)}_{R}$ and $\Gamma^{(2)}_{R}$ are engineered to be $\mathcal{O}(1)$ at large $c$ and so have good NR limits, which we refer to as $\Gamma^{(1)}$ and $\Gamma^{(2)}$. $\Gamma^{(1)}$ is manifestly gauge-invariant, but is not invariant under Milne boosts. $\Gamma^{(2)}$ is manifestly Milne-invariant, but not gauge-invariant. These connections are now straightforward to calculate,
\begin{align}
\begin{split}
(\Gamma^{(1)})^{\mu}{}_{\nu \rho} &= v^{\mu} \partial_{(\rho} n_{\nu)} + (\Gamma_{h})^{\mu}{}_{\nu \rho} +h^{\mu \sigma}
n_{(\nu} F_{\sigma) \rho} \,,
\\
(\Gamma^{(2)})^{\mu}{}_{\nu \rho}&=(\Gamma^{(1)})^{\mu}{}_{\nu \rho} +  h^{\mu \sigma} ( A_{\nu} \partial_{[\rho} n_{\sigma]} + A_{\rho} \partial_{[\nu} n_{\sigma]} )\,,
\end{split}
\end{align}
which we can easily recognize as the torsion-free parts of the gauge-invariant connection \eqref{giconnection} and the Milne invariant connection \eqref{miconnection} respectively. To obtain the torsionful connections, we can always further shift $\Gamma^{(1)}$ or $\Gamma^{(2)}$ by terms proportional to either $g^{\mu \sigma} \mathcal{F}_{\nu \rho} n_{\sigma}$ or $g^{\mu \sigma} {\cal F}^n_{\nu \rho} C_{\sigma}$ respectively, where the former preserves manifest gauge and the latter manifest Milne invariance.

%%%%%%%%%%%%%%%%%%%%%%%%%%%%%%%%%%%%%%%%%
\section{Magnetic moments}
\label{S:magnetic}
%%%%%%%%%%%%%%%%%%%%%%%%%%%%%%%%%%%%%%%%%

%%%%%%%%%%%%%%%%%%%%%%%%%%%%%%%%%%%%%%%%%
\subsection{The $g$-factor}
%%%%%%%%%%%%%%%%%%%%%%%%%%%%%%%%%%%%%%%%%

In two spatial dimensions, there is another term that can be introduced in the NR action for a free field which has a nice description in terms of NC geometry~\cite{Son:2013rqa,Geracie:2014nka}. In flat space it is
\beq
\label{gflat}
S_{g,flat} = \frac{g}{8} \int d^{2}x dt \, \epsilon^{ij} \, F_{ij} \, |\phi|^2 =
-i \frac{g}{4m} \int d^{2}x dt \, \epsilon^{ij} D_i \phi^* D_j \phi\,.
\eeq
Here $\epsilon^{ij}$ is the purely spatial epsilon tensor and the two equivalent forms of the term are related by integration by parts. This term is not invariant under the ``non-relativistic diffeomorphisms'' of~\cite{Son:2005rv}, but said transformation laws can be augmented by terms proportional to $g$ in such a way that the theory is invariant under them~\cite{Son:2013rqa}. In \cite{Jensen:2014aia} it was shown that these modified transformation laws can be completely reproduced and accounted for by a modified action of the Milne boost on $A_{\mu}$. All other fields retain their transformation properties in the presence of the $g$-term, in particular they still transform as standard tensors under reparameterizations. In order to make these statements manifest, we write \eqref{gflat} in a manifestly covariant form. There are two terms which reduce to~\eqref{gflat} in flat space,
\begin{align}
\begin{split}
S_{g_1} &= \frac{g_1}{8} \int d^{2}x dt\, \sqrt{\gamma} \, \epsilon^{\mu \nu \rho} \,n_{\mu} \, F_{\nu \rho} \, |\phi|^2\,,
 \\
\label{twog}
S_{g_2} &= - \frac{ig_2}{4m} \int d^{2}x dt\, \sqrt{\gamma} \, \epsilon^{\mu \nu \rho} \,n_{\mu} \, D_\nu \phi^* D_{\rho} \phi\,.
\end{split}
\end{align}
One interesting aspect of the covariant formulation \eqref{twog} is that these two terms are no longer equivalent. If we perform the integration by parts that gave us the two equivalent forms in~\eqref{gflat}, we pick up an extra term proportional to $\epsilon^{\mu\nu\rho}n_{\mu}F^n_{\nu\rho}|\phi|^2$. So in a general NC geometry, there are two independent magnetic moments, $g_1$ and $g_2$. In flat space only
\beq
g = g_1 + g_2\,.
\eeq
appears. Only the $g_2$ term was considered in \cite{Son:2013rqa,Geracie:2014nka,Jensen:2014aia}. It was shown in~\cite{Jensen:2014aia} that while $S_{g_2}$ alone is not Milne-invariant, the full NR action we obtain by adding the $g_2$ term to the free action from \eqref{freeaction} is Milne-invariant if we modify the Milne variation of $A_{\mu}$ to be
\beq
A_{\mu}\to A_{\mu} +   \psi_{\mu} - \frac{1}{2} n_{\mu} \psi^2 + n_{\mu} \frac{g_2}{4m}\epsilon^{\nu \rho \sigma} \partial_{\nu} \, (n_{\rho}\psi_{\sigma})\,.
\eeq
Since the new term in the variation is proportional to $n_{\mu}$, the only place where it contributes is the variation of $v^{\mu} A_{\mu}$; its coefficient is fixed to cancel the variation of the $g_2$ term. Instead, we can add the $g_1$ term and cancel its Milne transformation. It is easy to confirm that for
\beq
A_{\mu}\to A_{\mu} +\psi_{\mu} - \frac{1}{2} n_{\mu} \psi^2 + n_{\mu} \frac{g_1}{4m} \epsilon^{\nu \rho \sigma}
n_{\rho} \,  \partial_{\nu} \psi_{\sigma}\,,
\eeq
the action including the $g_1$ term is Milne invariant as long as we demand
\beq
\mathcal{B}^n \equiv \epsilon^{\mu \nu \rho} n_{\mu} \partial_{\nu} n_{\rho}   \label{ndn}  =0\,.
\eeq
If \eqref{ndn} is not satisfied, then the action of the Milne boost is more complicated and becomes non-analytic in background fields.\footnote{The full Milne boost when $\mathcal{B}^n$ is nonzero is
\beq
\label{nonlinear}
A_{\mu}\to A_{\mu}+ \psi_{\mu}- \frac{1}{2} n_{\mu} \psi^2 + n_{\mu} \frac{g_1}{4m} \frac{\epsilon^{\nu \rho \sigma}
n_{\rho} \,  \partial_{\nu} ( \psi_{\sigma} - \frac{1}{2} n_{\sigma} \psi^2)}{1+ \frac{g_1}{4m} \mathcal{B}^n}\, .
\eeq} The condition \eqref{ndn} has been argued in \cite{Geracie:2014nka} to be crucial in order to maintain NR causality. By the Frobenius theorem, any two points in a neighborhood in which $\mathcal{B}^n\neq 0$ can be reached by a curve whose tangent vector satisfies $t^{\mu}n_{\mu}>0$, and so this smells like a violation of causality. We will have more to say about this shortly. For now let us just note that having imposed \eqref{ndn}, we can add both magnetic moment terms to the action and get a Milne invariant theory as long as $A_{\mu}$ varies as
\beq
\label{modlaw}
A_{\mu} \to A_{\mu} +\psi_{\mu} - \frac{1}{2} n_{\mu} \psi^2 + n_{\mu} \frac{g_1}{4m} \epsilon^{\nu \rho \sigma}
n_{\rho} \,  \partial_{\nu}  \psi_{\sigma}
+ n_{\mu} \frac{g_2}{4m} \epsilon^{\nu \rho \sigma}
\partial_{\nu} \, (n_{\rho} \psi_{\sigma})\,.
\eeq

%%%%%%%%%%%%%%%%%%%%%%%%%%%%%%%%%%%%%%%%%
\subsection{Relativistic parent}
%%%%%%%%%%%%%%%%%%%%%%%%%%%%%%%%%%%%%%%%%

Instead of the two independent magnetic moments we can add in the NR theory, there is only one relativistic magnetic moment
\beq
\label{relmoment}
S_{g,R} = \frac{ig_R}{16  mc} \, \int d^{2}x dt \, \sqrt{-g} \, \epsilon^{\mu \nu \rho} {\cal F}_{\mu \nu}  \, (\Phi^* {\cal D}_{\rho} \Phi - \Phi {\cal D}_{\rho} \Phi^*) \,.
\eeq
This scalar magnetic moment only exists in $d=3$. Similar parity odd terms will exist in other dimensions, but their properties obviously depend on the dimension. The prefactor of $1/mc$ is needed to make sure that we get the correct NR magnetic moment terms at order 1; it follows on dimensional grounds.

There is also a new contribution at $\mathcal{O}(c^2)$ piece proportional to $\mathcal{B}^n$. As we mentioned above, \cite{Geracie:2014nka} stated that $\mathcal{B}^n=0$ is required for causality in the NR theory. It would be desirable to understand the status of this constraint in more detail. Exactly what goes wrong if $\mathcal{B}^n \neq 0$? For relativistic theories formulated on spacetimes with closed time-like curves, one can clearly see that the theory has intrinsic sicknesses. But non-zero $\mathcal{B}^n$ seems to be perfectly healthy in the relativistic parent. It also seems to be healthy for NR theories obtained by DLCQ ~\cite{Jensen:2014aia}. Any sickness in the NR theory at non-zero $\mathcal{B}^n$ would have to arise from the $c \rightarrow \infty$ limit.

Setting aside this important question, we return to magnetic moments. If we set $\mathcal{B}^n=0$, the NR magnetic moment terms as written arise naturally from a relativistic parent. If $\mathcal{B}^n\neq 0$, we get terms similar to the $g_1$ and $g_2$ terms above, but with extra powers of $\mathcal{B}^n$, as we discuss below. Since we get the NR magnetic moment terms as they stand at $\mathcal{B}^n=0$, we would like derive the modified transformation law \eqref{modlaw} from a NR limit of this relativistic parent even in this restricted case. Once we impose $\mathcal{B}^n=0$ we easily see that in the $c \rightarrow \infty$ limit our relativistic parent descends to a sum of $S_{g_1}$ and $S_{g_2}$ with
\beq
\label{relg}
g_1 = 2 g_R\,, \quad \quad g_2 = -g_R\,.
\eeq
We should however note that even after imposing $\mathcal{B}^n=0$ our troubles are not over. One of the main motivations for starting out with the relativistic parent is that it automatically generates Milne-invariant actions. All relativistic fields are manifestly Milne invariant. However $n_{\mu}$, and hence $\mathcal{B}^n$, and $b_{\mu}$ are not. They are just the leading $\mathcal{O}(c^2)$ pieces in the metric and gauge field. We can set to zero the term proportional to $\mathcal{B}^n$ in the action simply by postulating that the constraint \eqref{ndn} be obeyed. This does, however, not ensure that its Milne variation vanishes. The $\mathcal{B}^n$ term in the relativistic action will generate a new contribution to the Milne variation to the action even when we set it to zero in the end. There is one easy way to deal with this: we keep $\mathcal{B}^n$ nonzero when determining the Milne variation of all fields and only set $\mathcal{B}^n=0$ in the very end. In fact, it is easy to see that the $\mathcal{O}(c^2)$ contribution the magnetic moment term makes to the action can even be cancelled at non-zero $\mathcal{B}^n$ by modifying the relation \eqref{nbrel} between $n_{\mu}$ and $b_{\mu}$. In this case we can still take a consistent NR limit, that is the action has a good large $c$ limit. It just doesn't reduce to the simple magnetic moment terms $S_{g_1}$ and $S_{g_2}$. We will derive the Milne transformations of $A_{\mu}$ by studying this theory at non-zero $\mathcal{B}^n$ and then, in the end, will set $\mathcal{B}^n=0$ to get the simple transformation rules \eqref{modlaw} for the theory with $S_{g_1}$ and $S_{g_2}$. But before we do so, we briefly want to discuss why there is a single relativistic parent term even though there are two non-trivial NR terms.

%%%%%%%%%%%%%%%%%%%%%%%%%%%%%%%%%%%%%%%%%
\subsection{A note on gauge invariance}
%%%%%%%%%%%%%%%%%%%%%%%%%%%%%%%%%%%%%%%%%

Eq.~\eqref{relmoment} is the unique gauge-invariant relativistic magnetic moment we can add to the scalar action. How then can we understand that in the NR limit we can have two independent terms? Shouldn't each NR term have an individual parent? We can answer this question by focusing on the putative parent of the $g_1$ term. We can write
\beq
S \sim  \int d^2x dt \sqrt{-g} \, \epsilon^{\mu \nu \rho} C_{\mu} {\cal F}_{\nu \rho} |\Phi|^2\,,
\eeq
which reduces to $S_{g_1}$ in the large $c$ limit, provided we set the leading $\mathcal{B}^n =0$. Note that $C \wedge {\cal F}$ is the standard 2+1 dimensional Chern-Simons (CS) term, which is of course gauge-invariant on its own right. Even though $C$ appears explicitly, the gauge variation of the CS term is a total derivative and so its integral is gauge-invariant. This however fails in the magnetic moment above. Integrating by parts will leave a non-trivial variation of the action proportional to derivatives of $|\Phi|^2$. Note however that if we replace $C_{\mu}$ with $c^2 b_{\mu}$, its leading part, the action suddenly is gauge-invariant in the NR sense.

The reason for this is that NR gauge-invariance is a much weaker requirement than the full relativistic gauge-invariance. In the relativistic theory we can expand the gauge parameter itself in a power series in $c$:
\beq
\Lambda = \Lambda_0 c^2 + \Lambda_1 + \ldots \,.
\eeq
Relativistic gauge invariance requires that we are invariant under all gauge transformations, including those parametrized by $\Lambda_0$. Performing a $\Lambda_0$ gauge transformation however is not consistent with the NR limit. The leading piece in the gauge field is tied by \eqref{nbrel} to the metric. In the NR theory, $\mathcal{O}(c^2)$ gauge transformations are no longer allowed and the standard NR gauge transformations are generated by $\Lambda_1$. The $S_{g_1}$ term in the action is perfectly gauge invariant under $\mathcal{O}(1)$ gauge transformations in the large $c$ limit, and so allowed in the NR theory. It is however forbidden in the relativistic parent.

This is a very important lesson to draw from this simple example. Our procedure of obtaining NR actions and transformation laws from relativistic parents gives a very physical way of interpreting the NC data. It gives a natural way to understand the role of Milne boosts. It automatically generates Milne and gauge-invariant actions. It does, however, not give the most general NR terms allowed. Requiring relativistic gauge invariance is a stronger constraint than NR gauge invariance.

%%%%%%%%%%%%%%%%%%%%%%%%%%%%%%%%%%%%%%%%%
\subsection{The modified Milne boost}
%%%%%%%%%%%%%%%%%%%%%%%%%%%%%%%%%%%%%%%%%

Without any further ado, let us now proceed to deriving the modified Milne boost \eqref{modlaw} for the special case $g_1=2g_R, g_2=-g_R$, from the relativistic parent theory. As we discussed above, we need to work at non-zero $\mathcal{B}^n$ for now and only will, in the very end, set $\mathcal{B}^n=0$. In order to cancel the $\mathcal{O}(c^2)$ terms in the action, we need to modify the relation \eqref{nbrel} between $n_{\mu}$ and $b_{\mu}$. Avoiding $\mathcal{O}(c^4)$ terms still requires $h^{\mu \nu} b_{\mu} b_{\nu}=0$, which implies $b_{\mu} =  n_{\mu}/\alpha$. Demanding that the $\mathcal{O}(c^2)$ terms also vanishes fixes
\beq
\label{alpha} \alpha = \sqrt{  1 + \frac{g_R}{2m} \mathcal{B}^n}\,.
\eeq
Note that once we set $\mathcal{B}^n=0$ we will be back to $\alpha=1$, but to derive the Milne transformation rules of the theory with a magnetic moment we need to track the non-trivial factors of $\alpha$ and only set it to 1 in the end. Plugging $b_{\mu}=n_{\mu}/\alpha$ with \eqref{alpha} back into the relativistic magnetic moment action and then taking $c\to\infty$ does give us a consistent NR magnetic moment that is Milne-invariant for any value of $\mathcal{B}^n$ (just like the $g_2$ term by itself was as well), but this theory will have factors of $\alpha$ in it and with it non-analytic dependence\footnote{For the sake of concreteness let us spell out the action at non-zero $\mathcal{B}^n$:
\beq
\label{fromparent}
S = \int d^{2}x dt\,\sqrt{\gamma}\left\{  \frac{iv^{\mu}}{2 \alpha} (\phi^* D_{\mu} \phi - \phi D_{\mu} \phi^*)  - \frac{h^{\mu \nu}}{2m} D_{\mu} \phi^* D_{\nu} \phi +\frac{g_R}{4m}  \, \epsilon^{\mu \nu \rho} \, \frac{n_{\mu}}{\alpha} \, \left ( m\, F_{\nu \rho} \, |\phi|^2 + i
 \, D_\nu \phi^* D_{\rho} \phi \right ) \right\}\,.
\eeq
Note that this action is an equally valid NR magnetic moment-like term that has all the same symmetries as the original $g_1$ and $g_2$ terms. It involves, however, an infinite number of higher derivative terms hiding in $\alpha$ and so, from the point of view of low energy effective theory, looks rather unnatural.} on $\mathcal{B}^n$.

As in \eqref{milne} we now can simply read of the transformation of the various fields under the Milne boost. The transformations of $h^{\mu \nu}$, $v^{\mu}$ and $n_{\mu}$ are as before, but for $b_{\mu}$ we have
\beq
\delta b_{\mu} =   \frac{\delta n_{\mu}}{\alpha} -  \frac{n_{\mu}}{\alpha^2} \delta \alpha \,.
\eeq
Evaluating these variations and then setting $\alpha=1$ (that is $\mathcal{B}^n=0$) we get
\beq
c^2 b_{\mu} \to c^2  b_{\mu}  -  \Psi_{\mu} + \frac{g_R}{4 m} n_{\mu} \epsilon^{\nu \rho \sigma} \left [ \Psi_{\nu} \partial_{\rho} n_{\sigma} + n_{\nu} \partial_{\rho} \Psi_{\sigma} \right ]+ \mathcal{O}(c^{-2})\,.
\eeq
This change has to be compensated, as before, by $A_{\mu} \to A_{\mu} - c^2 \delta b_{\mu}$.
Recalling that $\Psi_{\mu}$ was constrained to take the form \eqref{Psi} to ensure that $h_{\mu \nu}$ remains degenerate this finally yields\footnote{It is also straightforward to keep track of non-zero $\mathcal{B}^n$ in this expression. In this case we would get
\beq
\label{finan}
A_{\mu} \to A_{\mu} + \frac{\Psi_{\mu}}{\alpha} - \frac{g_R}{4 m \alpha^2} n_{\mu} \epsilon^{\nu \rho \sigma} \left [ \Psi_{\nu} \partial_{\rho}\left(  \frac{n_{\sigma}}{\alpha}\right) + \frac{n_{\nu}}{\alpha} \partial_{\rho} \Psi_{\sigma} \right ].
\eeq
This is similar to the transformation law we would get in the NR theory with $g_1=2g_R, g_2=-g_R,$ with the full non-linear transformation law of the previous footnote \eqref{nonlinear}. The fact that the appearances of $\alpha$ do not quite match is due to the fact that already on the level of the action what we get from the relativistic parent, \eqref{fromparent}, is really not the same as the NR magnetic moment terms from \eqref{twog} but differs by factors of $\alpha$, the higher derivative modification we pointed out in the previous footnote.
}
\beq
\label{finan}
\delta A_{\mu} = P^{\nu}_{\mu} \psi_{\nu} - \frac{1}{2} n_{\mu} \psi^2 - \frac{g_R}{4 m} n_{\mu} \epsilon^{\nu \rho \sigma} \left [ \psi_{\nu} \partial_{\rho} n_{\sigma} + n_{\nu} \partial_{\rho} \psi_{\sigma} \right ]\,.
\eeq
Here we used that $\mathcal{B}^n=0$, which allows us to simply replace $\Psi_{\mu}$ with $ \psi_{\mu}$ in both of the terms contracted with $\epsilon^{\nu \rho \sigma}$. Amazingly this final answer \eqref{finan} agrees exactly with the NR rules for the modified Milne boosts in the presences of magnetic moment terms, \eqref{modlaw}, when we specialize it to the case of $g_1=2g_R$, $g_2=-g_R$ that we inherit from our limit, see \eqref{relg}.

%%%%%%%%%%%%%%%%%%%%%%%%%%%%%%%%%%%%%%%%%
\section{Hydrodynamics}
\label{S:hydro}
%%%%%%%%%%%%%%%%%%%%%%%%%%%%%%%%%%%%%%%%%

The second setting where we revisit the non-relativistic limit is hydrodynamics. Hydrodynamics is a low-energy, long-wavelength effective description of field theory at nonzero temperature, encoding the dynamics of the relaxation of conserved quantities. See~\cite{Kovtun:2012rj} for an excellent review.

The basic ingredients of relativistic hydrodynamics with a global $U(1)$ symmetry are:
\begin{enumerate}
\item Thermal equilibria in flat space are specified by a temperature $T$, chemical potential $\mu_r$, and normalized velocity $\mathcal{U}^{\mu}$ satisfying $\mathcal{U}^2 = -c^2$. In hydrodynamics one promotes these parameters to classical fields. The $(T,\mu_R,\mathcal{U}^{\mu})$ are the fluid variables in the hydrodynamic description.

\item Unlike Wilsonian effective field theory, one continues by specifying the one-point functions of the stress tensor $t^{\mu\nu}$ and $U(1)$ current $j^{\mu}$ in terms of the fluid variables and the background fields $(g_{\mu\nu},C_{\mu})$. One does so in a gradient expansion, wherein the background fields are taken to be $\mathcal{O}(\partial^0)$. The term $n^{th}$ order hydrodynamics refers to fluid mechanics where the constitutive relations have been specified to $\mathcal{O}(\partial^n)$.

\item Enforce the Ward identities
\beq
\label{E:Rward}
\mathcal{D}_{\nu}t^{\mu\nu} = \mathcal{F}^{\mu}{}_{\nu}j^{\nu} \,, \qquad \mathcal{D}_{\mu}j^{\mu} = 0\,,
\eeq
as equations of motion which fix the fluid variables $(T,\mu_R,\mathcal{U}^{\mu})$.

\item Demand a local version of the second Law of thermodynamics. More precisely, one demands the existence of an entropy current $S^{\mu}$ whose divergence is non-negative for physical fluid flows, i.e. those which satisfy~\eqref{E:Rward}. In Subsection~\ref{S:hydroPrelim} we show how the non-relativistic fluid variables come from the relativistic ones,

\end{enumerate}
A similar itemized list exists for non-relativistic hydrodynamics, although it takes some work to ensure invariance under Galilean boosts. The fluid variables of the non-relativistic fluid mechanics are a local temperature $T$, chemical potential $\mu$ for particle number, and a Milne boost-invariant velocity $u^{\mu}$ satisfying $u^{\mu}n_{\mu}=1$. See~\cite{Jensen:2014ama} for the details.

Below, we take the non-relativistic limit of relativistic hydrodynamics in the same way as we did for perturbative scalar field theory in the previous two Sections. In that case, we directly had the effective action and so it was relatively easy to manifest the regularity of the $c\to\infty$ limit. Hydrodynamics is slightly more complicated insofar as we work with the one-point functions rather than the partition function itself. This complication has led to some confusion in the literature which we resolve. The essential point is that a certain combination of $\mathcal{O}(c^{-1})$ terms in $t^{\mu\nu}$ and $j^{\mu}$ must vanish. See Subsection~\ref{S:NRonePointFromLimit} for details.

We will show how the large $c$ limit maps each of the items above into the corresponding item for non-relativistic hydrodynamics. The non-relativistic fluid variables emerge from the relativistic ones in Subsection~\ref{S:hydroPrelim}. We obtain the non-relativistic stress tensor and Milne boost-invariant energy current from the relativistic stress tensor and current in Subsection~\ref{S:NRonePointFromLimit}. The relativistic Ward identities imply the non-relativistic ones, and the relativistic entropy condition implies the non-relativistic one, as we show in Subsections~\ref{S:NRwardFromLimit} and~\ref{S:entropy}.

%%%%%%%%%%%%%%%%%%%%%%%%%%%%%%%%%%%%%%%%%
\subsection{Some preliminaries}
\label{S:hydroPrelim}
%%%%%%%%%%%%%%%%%%%%%%%%%%%%%%%%%%%%%%%%%

To proceed efficiently we need a few basic results.

The relativistic stress tensor and $U(1)$ current are defined by variations of the generating functional as
\beq
\label{E:Rcurrents}
\delta W_R = \int d^dx \sqrt{-g} \left\{ \delta C_{\mu} j^{\mu} + \frac{1}{2}\delta g_{\mu\nu} t^{\mu\nu}\right\}\,.
\eeq
We consider relativistic fluid mechanics coupled to the same background spacetime and gauge field we considered in Section~\ref{S:RtoNR},
\begin{align}
\begin{split}
\label{E:Rbackground}
g_{\mu\nu} &= - c^2 n_{\mu}n_{\nu} + h_{\mu\nu}\,,
\\
C_{\mu} & = m c^2 n_{\mu} + mA_{\mu}\,.
\end{split}
\end{align}
Correspondingly, the relativistic chemical potential is
\beq
\mu_R = m c^2 + m\mu\,,
\eeq
where $\mu$ is regular as $c\to \infty$. We also separate the relativistic fluid velocity into components which are longitudinal and transverse to $n_{\mu}$,
\beq
\mathcal{U}^{\mu} =\gamma\, u^{\mu} =  \gamma (v^{\mu} + w^{\mu})\,, \qquad \gamma = \left( 1 - \frac{w^2}{c^2}\right)^{-\frac{1}{2}}\,,
\eeq
where $n_{\mu}w^{\nu}=0$. Here, $\gamma$ is the usual relativistic kinematic factor ensuring that $\mathcal{U}^2 = - c^2$. It is not the determinant of $\gamma_{\mu\nu}$, but we hope this is clear from the context. It then follows that $u^{\mu}n_{\mu}=1$. We consider fluid flows for which $u^{\mu}$ and $T$ are regular as $c\to\infty$. Ultimately, $(T,\mu,u^{\mu})$ will become the fluid variables of the non-relativistic hydrodynamics we find in the $c\to\infty$ limit. We will also use that
\beq
\mathcal{U}_{\mu} = \gamma\left( - c^2 n_{\mu} + w_{\mu}\right)\,,
\eeq
where $w_{\mu}=h_{\mu\nu}w^{\nu}$.

Recall from Section~\ref{S:RtoNR} that the non-relativistic Milne boosts come from a redundancy in the relativistic description. The boosts are $\mathcal{O}(c^{-2})$ redefinitions of $n_{\mu}$, compensated by $\mathcal{O}(1)$ redefinitions of $h_{\mu\nu}$ and $A_{\mu}$ in such a way as to keep $g_{\mu\nu}$ and $C_{\mu}$ in~\eqref{E:Rbackground} fixed. The relativistic fluid velocity $\mathcal{U}^{\mu}$, and so $u^{\mu}$, are invariant under any such redundancy, which gives that the $u^{\mu}$ defined here is Milne-invariant.

In Galilean theories coupled to Newton-Cartan geometry, there is a spatial stress tensor $T_{\mu\nu}$, a momentum current $\mathcal{P}_{\mu}$, a number current $J^{\mu}$, and an energy current $\mathcal{E}^{\mu}$. Taken together, these currents define a sort of ``stress tensor complex'' for the non-relativistic theory, which are described with constitutive relations in non-relativistic hydrodynamics. They are defined by functional variation~\cite{Geracie:2014nka,Jensen:2014aia}
\beq
\label{E:NRcurrents}
\delta W = \int d^dx \sqrt{\gamma} \left\{ \delta A_{\mu} J^{\mu} - \delta \bar{v}^{\mu}\mathcal{P}_{\mu} - \delta n_{\mu} \mathcal{E}^{\mu} - \frac{1}{2}\delta \bar{h}^{\mu\nu}T_{\mu\nu}\right\}\,.
\eeq
Here we have let $W$ depend on the overcomplete set of background fields $(n_{\mu},v^{\mu},h^{\mu\nu},A_{\mu})$ (recall that $n_{\mu}$ is algebraically determined by $v^{\mu}$ and $h^{\mu\nu}$), and taken variations in such a way as to keep $n_{\mu}v^{\mu}=1$ and $h^{\mu\nu}n_{\nu}=1$. We work in a convention where the variations of $n_{\mu}$ are arbitrary, so that the variations of $v^{\mu}$ and $h^{\mu\nu}$ are partially fixed as
\begin{align}
\begin{split}
\label{E:constrainedVars}
\delta v^{\mu} &= - v^{\mu}v^{\nu}\delta n_{\nu} + P^{\mu}_{\nu} \delta \bar{v}^{\nu}\,,
\\
\delta h^{\mu\nu} & = - \left( v^{\mu}h^{\nu\rho} + v^{\nu}h^{\mu\rho}\right)\delta n_{\rho} + P^{\mu}_{\rho}P^{\nu}_{\sigma} \delta \bar{h}^{\rho\sigma}\,,
\end{split}
\end{align}
where $\delta \bar{v}^{\mu}$ and $\delta \bar{h}^{\mu\nu}$ are arbitrary. Note that due to~\eqref{E:constrainedVars}, $\mathcal{P}_{\mu}$ and $T_{\mu\nu}$ are spatial insofar as $\mathcal{P}_{\mu}v^{\mu}$ and $T_{\mu\nu}v^{\nu}$ both vanish.

The symmetries of the problem yield Ward identities~\cite{Geracie:2014nka,Jensen:2014aia} for the stress tensor and currents: the Milne Ward identity equates momentum and particle number currents,
\beq
\label{E:milneWard}
\mathcal{P}_{\mu} =  h_{\mu\nu}J^{\nu}\,,
\eeq
while $U(1)$ gauge and coordinate reparameterization invariance imply conservation equations. These are most concisely written in terms of Milne boost-invariant data. These include a spacetime stress tensor $\mathcal{T}^{\mu\nu}$~\cite{Jensen:2014aia} and energy current $\tilde{\mathcal{E}}^{\mu}$~\cite{Jensen:2014ama}
\begin{align}
\begin{split}
\label{E:TandE}
\mathcal{T}^{\mu\nu} &= T^{\mu\nu} + v^{\mu}\mathcal{P}^{\nu} + v^{\nu}\mathcal{P}^{\mu} +  v^{\mu}v^{\nu}n_{\rho}J^{\rho}\,,
\\
\tilde{\mathcal{E}}^{\mu}& = \mathcal{E}^{\mu} - \left( u_{\nu}-\frac{1}{2}n_{\nu}u^2\right)\mathcal{T}^{\mu\nu}\,,
\end{split}
\end{align}
where indices are raised with $h^{\mu\nu}$, $u_{\mu}=h_{\mu\nu}u^{\nu}$, and $u^2 = u_{\mu}u^{\mu}$. Note that
\beq
\mathcal{T}^{\mu\nu}n_{\nu} = J^{\mu}\,,
\eeq
by virtue of~\eqref{E:milneWard}. We also use the fluid velocity to define a Milne boost-invariant version of $h_{\mu\nu}$,
\beq
\tilde{h}_{\mu\nu} = h_{\mu\nu} - \left( n_{\mu}u_{\nu}+n_{\nu}u_{\mu}\right) + n_{\mu}n_{\nu}u^2\,,
\eeq
for which $\tilde{h}_{\mu\nu}u^{\nu}=0$, and a new $U(1)$ and gravitational connection~\cite{Jensen:2014ama}
\begin{align}
\begin{split}
\label{E:milneD}
\tilde{\Gamma}^{\mu}{}_{\nu\rho} & = u^{\mu}\partial_{\rho}n_{\nu} + \frac{h^{\mu\sigma}}{2}\left( \partial_{\nu}\tilde{h}_{\rho\sigma} + \partial_{\rho}\tilde{h}_{\nu\sigma} - \partial_{\sigma}\tilde{h}_{\nu\rho}\right) +  h^{\mu\sigma} n_{(\nu}F_{\rho)\sigma}\,,
\\
\tilde{A}_{\mu} & = A_{\mu} +  u_{\mu} - \frac{1}{2}n_{\mu}u^2\,.
\end{split}
\end{align}
We denote the corresponding covariant derivative as $\tilde{D}_{\mu}$. In terms of it and $\tilde{\mathcal{G}}_{\mu} = - F^n_{\mu\nu}u^{\nu} = - E^n_{\mu}$, the Ward identities are then~\cite{Jensen:2014ama}
\begin{align}
\begin{split}
\label{E:NRward}
\left( \tilde{D}_{\mu} - \tilde{\mathcal{G}}_{\mu}\right) \tilde{\mathcal{E}}^{\mu} &= \mathcal{G}_{\mu}\mathcal{E}^{\mu} - \tilde{h}_{\rho(\mu}\tilde{D}_{\nu)}u^{\rho} \tilde{T}^{\mu\nu}\,,
\\
\left( \tilde{D}_{\nu}-\tilde{\mathcal{G}}_{\nu}\right)\mathcal{T}^{\mu\nu} & = - (F^n)^{\mu}{}_{\nu} \tilde{\mathcal{E}}^{\nu}\,.
\end{split}
\end{align}
The appearance of $\tilde{\mathcal{G}}_{\mu}$ in the covariant divergences is just due to the torsion, and it is easy to see that, for vector fields, the covariant divergence is just the usual one
\beq
\left( \tilde{D}_{\mu}-\tilde{\mathcal{G}}_{\mu}\right) \mathfrak{v}^{\mu} = \frac{1}{\sqrt{\gamma}}\partial_{\mu}\left( \sqrt{\gamma}\mathfrak{v}^{\mu}\right)\,,
\eeq
with a volume element $\sqrt{\gamma}$.  The longitudinal component of the stress tensor Ward identity is just the conservation of number,
\beq
\label{E:numberWard}
\left(\tilde{D}_{\mu}-\tilde{\mathcal{G}}_{\mu}\right)J^{\mu} = 0\,.
\eeq

%%%%%%%%%%%%%%%%%%%%%%%%%%%%%%%%%%%%%%%%%
\subsection{The large $c$ limit}
%%%%%%%%%%%%%%%%%%%%%%%%%%%%%%%%%%%%%%%%%

\subsubsection{The non-relativistic stress tensor and energy current}
\label{S:NRonePointFromLimit}

Now let us relate the one-point functions of the relativistic theory to those of the NR theory attained in the large $c$ limit. As usual, we assume that the relativistic $W_R$ evaluated for the background fields~\eqref{E:Rbackground} is regular as $c\to\infty$,
\beq
\lim_{c\to\infty}W_R[g_{\mu\nu}=- c^2n_{\mu}n_{\nu}+h_{\mu\nu},C_{\mu}=mc^2 n_{\mu}+mA_{\mu}]  = W[n_{\mu},h_{\mu\nu},A_{\mu}]\,.
\eeq
Putting the definition of the relativistic~\eqref{E:Rcurrents} and non-relativistic currents~\eqref{E:NRcurrents} and the relativistic background fields~\eqref{E:Rbackground}, we see that
\begin{align}
\begin{split}
\label{E:NRnumberFromLimit}
J^{\mu} &= \frac{1}{\sqrt{\gamma}}\frac{\delta W}{\delta A_{\mu}}=\lim_{c\to\infty} \left\{ \left( \frac{\sqrt{-g}}{\sqrt{\gamma}}\right)\frac{m}{\sqrt{-g}}\frac{\delta W_R}{\delta C_{\mu}}\right\}
\\
& =m\lim_{c\to\infty} c\, j^{\mu}\,,
\end{split}
\end{align}
where we have used
\beq
\sqrt{-g} = c \sqrt{\gamma}\,, \qquad \frac{\delta C_{\mu}}{\delta A_{\nu}} = m  \delta^{\nu}_{\mu}\,.
\eeq
The mighty chain rule also gives
\begin{align}
\begin{split}
\label{E:NRcurrentsFromLimit}
\mathcal{P}_{\mu} & = \lim_{c\to\infty} c\, h_{\mu\nu}t^{\nu\rho}n_{\rho}\,,
\\
T_{\mu\nu} & = \lim_{c\to\infty} c\, h_{\mu\rho}h_{\nu\sigma}t^{\rho\sigma}\,,
\\
\mathcal{E}^{\mu} & = \lim_{c\to\infty} c^3\left( t^{\mu\nu}n_{\nu} - m j^{\mu}\right)\,.
\end{split}
\end{align}
In general, the relativistic stress tensor and current are $\mathcal{O}(c^{-1})$ at large $c$. To get a well-defined energy current $\mathcal{E}^{\mu}$ from the large $c$ limit, we obviously need that the $\mathcal{O}(c^{-1})$ term in $t^{\mu\nu} n_{\nu}-m j^{\mu}$ vanishes,
\beq
\label{E:aCaseOfTheObvious}
0  = \lim_{c\to\infty} c\left( t^{\mu\nu}n_{\nu} - mj^{\mu}\right)\,,
\eeq
Note that this automatically implies the Milne Ward identity,
\beq
0 = h_{\mu\nu}\left( \lim_{c\to\infty} c\left\{ t^{\nu\rho}n_{\rho}-m j^{\nu}\right\}\right) = \mathcal{P}_{\mu}-h_{\mu\nu}J^{\nu}\,.
\eeq
A proper discussion of hydrodynamic field redefinitions (as in~\cite{Jensen:2014ama}) is beyond the scope of what we do here, however, we observe that the RHS of~\eqref{E:aCaseOfTheObvious} is a combination of the constitutive relations which is invariant under redefinitions of $(T,\mu_R,\mathcal{U}^{\mu})$ which are regular as $c\to\infty$.

This is essentially the only non-trivial component in the large $c$ limit of relativistic hydrodynamics. Regularity of the limit requires~\eqref{E:aCaseOfTheObvious}, which in turn enforces the Milne Ward identity on the resulting non-relativistic constitutive relations.\footnote{The standard large $c$ limit of relativistic hydrodynamics, nicely reviewed in Kaminski and Moroz~\cite{Kaminski:2013gca}, is in the same spirit as ours. Applied to viscous hydrodynamics in flat space, that limit indeed satisfies~\eqref{E:aCaseOfTheObvious}. The authors of~\cite{Kaminski:2013gca} also take the large $c$ limit of parity-violating first-order hydrodynamics in two spatial dimensions~\cite{Jensen:2011xb}. However, upon converting the conventions of~\cite{Kaminski:2013gca} to ours, their large $c$ limit of the parity-violating hydrodynamics does not satisfy~\eqref{E:aCaseOfTheObvious}. That is, Kaminski and Moroz scale one of the parity-violating response coefficients with $c$ in such a way that the $c\to\infty$ limit is not regular. This explains the result obtained in the Appendix of~\cite{Jensen:2014ama} that the parity-violating hydrodynamics obtained by Kaminski and Moroz is inconsistent with Galilean boost invariance.}

Comparing~\eqref{E:NRnumberFromLimit} and~\eqref{E:NRcurrentsFromLimit} with~\eqref{E:TandE}, we see that the spacetime stress tensor $\mathcal{T}^{\mu\nu}$ is just the limit of the relativistic stress tensor,
\beq
\label{E:calTfromLimit}
\mathcal{T}^{\mu\nu} = \lim_{c\to\infty} c\, t^{\mu\nu}\,.
\eeq
What of the Milne boost-invariant energy current in~\eqref{E:TandE}? Using that
\beq
\lim_{c\to\infty} (\mathcal{U}_{\mu} + c^2n_{\mu}) = u_{\mu} - \frac{1}{2}n_{\mu}u^2\,,
\eeq
we find
\beq
\label{E:calEfromLimit}
\tilde{\mathcal{E}}^{\mu} = - \lim_{c\to\infty} c\left( t^{\mu\nu}\mathcal{U}_{\nu} + mc^2 j^{\mu}\right)\,.
\eeq

\subsubsection{The non-relativistic velocity}

By assumption, we study fluid flows for which the relativistic fluid velocity $\mathcal{U}^{\mu}=\gamma u^{\mu}$ is regular at large $c$,
\begin{equation*}
\lim_{c\to\infty} \mathcal{U}^{\mu} = u^{\mu}\,.
\end{equation*}
What about the derivative of $\mathcal{U}^{\mu}$? Here, we digress on the relation between $\mathcal{D}_{\mu}\mathcal{U}_{\nu}$ and $\tilde{D}_{\mu}u^{\nu}$. This will prove necessary when we recover the energy Ward identity~\eqref{E:NRward} from the relativistic Ward identities in the next Subsubsection.

We begin by defining the relativistic projector
\beq
\Delta_{\mu\nu} = g_{\mu\nu} + \frac{\mathcal{U}_{\mu}\mathcal{U}_{\nu}}{c^2}\,,
\eeq
which satisfies $\Delta_{\mu\nu}\mathcal{U}^{\nu} = 0$ and $\Delta_{\mu}^{\rho}\Delta_{\rho}^{\nu} = \Delta_{\mu}^{\nu}$. $\Delta$ has a regular large $c$ limit,
\beq
\label{E:DeltaLimit}
\lim_{c\to\infty} \Delta_{\mu\nu} = \tilde{h}_{\mu\nu}\,, \qquad \lim_{c\to\infty} \Delta_{\mu}^{\nu} = \tilde{P}_{\mu}^{\nu} = \tilde{h}_{\mu\rho}h^{\nu\rho}\,, \qquad \lim_{c\to\infty} \Delta^{\mu\nu} = h^{\mu\nu}\,.
\eeq
In terms of $\Delta_{\mu\nu}$, the symmetric part of $\mathcal{D}_{\mu}\mathcal{U}_{\nu}$ is
\begin{subequations}
\label{E:DUR}
\begin{align}
\mathcal{D}_{(\mu}\mathcal{U}_{\nu)} &= \frac{\Delta_{\mu\nu}}{d-1}\vartheta^R + \sigma^R_{\mu\nu} -\mathcal{U}_{(\mu} a^R_{\nu)}\,, & \vartheta^R &= \mathcal{D}_{\mu}\mathcal{U}^{\mu}\,,  & a^R_{\mu}&=\frac{1}{c^2} \mathcal{U}^{\nu}\mathcal{D}_{\nu}\mathcal{U}_{\mu}\,,
\end{align}
and
\beq
 (\sigma^R)^{\mu\nu} = \Delta^{\mu\rho}\Delta^{\nu\sigma}\left( \mathcal{D}_{(\rho}\mathcal{U}_{\sigma)}  - \frac{\vartheta^R}{d-1}\Delta_{\rho\sigma}\right) =\frac{\Delta^{\mu\rho}\Delta^{\nu\sigma}}{2}\pounds_{\mathcal{U}} \Delta_{\rho\sigma} - \frac{\vartheta^R}{d-1}\Delta^{\mu\nu}\,,
\eeq
\end{subequations}
where $\vartheta^R, \sigma^R$ and $a^R$ are the relativistic expansion, shear and acceleration. The non-relativistic velocity $u^{\mu}$ has a similar decomposition of its derivative~\cite{Jensen:2014ama},
\begin{subequations}
\begin{align}
\label{E:DuNR}
\tilde{D}_{\mu}u^{\nu} & = - n_{\mu} E^{\nu} + \frac{1}{2}B_{\mu}{}^{\nu} +\tilde{h}_{\mu\rho}\sigma^{\nu\rho} + \frac{\vartheta}{d-1}h^{\mu\nu}\,, & \vartheta & = \tilde{D}_{\mu}u^{\mu}\,,
\\
 E_{\mu} & = \tilde{F}_{\mu\nu}u^{\nu}\,, &B_{\mu\nu} & = \tilde{P}_{\mu}^{\rho}\tilde{P}_{\nu}^{\sigma}\tilde{F}_{\rho\sigma}\,.
\end{align}
and
\beq
\sigma^{\mu\nu} = \frac{1}{2}\left( h^{\mu\rho}\tilde{D}_{\rho}u^{\nu} + h^{\nu\rho}\tilde{D}_{\rho}u^{\mu} - \frac{2\vartheta}{d-1}h^{\mu\nu}\right) = \frac{h^{\mu\rho}h^{\nu\sigma}}{2}\pounds_u \tilde{h}_{\rho\sigma}- \frac{\vartheta}{d-1}h^{\mu\nu}\,.
\eeq
\end{subequations}
Using that
\begin{equation*}
\vartheta^R = \frac{1}{\sqrt{-g}}\partial_{\mu}(\sqrt{-g}\mathcal{U}^{\mu})\,, \qquad \vartheta = \frac{1}{\sqrt{\gamma}}\partial_{\mu}(\sqrt{\gamma}u^{\mu})\,,
\end{equation*}
along with the expressions for $\sigma^R$ and $\sigma$ in terms of Lie derivatives, it immediately follows that the relativistic expansion and shear just become the non-relativistic expansion and shear
\beq
\label{E:limitTheta}
\lim_{c\to\infty} \vartheta^R = \vartheta\,, \qquad \lim_{c\to\infty} (\sigma^R)^{\mu\nu} = \sigma^{\mu\nu}\,,
\eeq

There are other one-derivative tensor structures which frequently appear in hydrodynamics. The background electromagnetic field and anti-symmetric derivative of $\mathcal{U}_{\mu}$ have large $\mathcal{O}(c^2)$ pieces,
\beq
\lim_{c\to\infty} \frac{\mathcal{F}_{\mu\nu}}{c^2} = mF^n_{\mu\nu}\,, \qquad \lim_{c\to\infty} \frac{\partial_{\mu}\mathcal{U}_{\nu}-\partial_{\nu}\mathcal{U}_{\mu}}{c^2} = - F^n_{\mu\nu}\,.
\eeq
This implies that the rest-frame electric field $\mathcal{F}_{\mu\nu}\mathcal{U}^{\nu}$ and acceleration $a_{\mu}$ obey\beq
\label{E:limitAccel}
\lim_{c\to\infty} \frac{\mathcal{F}_{\mu\nu}\mathcal{U}^{\nu}}{c^2} = mE^n_{\mu}\,, \qquad \lim_{c\to\infty}a_{\mu} = E^n_{\mu}\,.
\eeq
Similarly, the magnetic field $\Delta_{\mu}^{\rho}\Delta_{\nu}^{\sigma}\mathcal{F}_{\rho\sigma}$ and vorticity $\omega_{\mu\nu} = \Delta_{\mu}^{\rho}\Delta_{\mu}^{\rho}(\partial_{\rho}\mathcal{U}_{\sigma}-\partial_{\sigma}\mathcal{U}_{\rho})$ give the magnetic part of $F^n$ in the $c\to\infty$ limit.

Another tensor structure that frequently appears in hydrodynamics is $\mathcal{F}^{\mu\nu}\mathcal{U}_{\nu} - T \Delta^{\mu\nu}\partial_{\nu}\left( \frac{\mu_R}{T}\right)$. It is also $\mathcal{O}(c^2)$ at large c, and its limit is
\beq
\label{E:limitV}
\lim_{c\to\infty} \frac{1}{c^2}\left\{\mathcal{F}^{\mu\nu}\mathcal{U}_{\nu} - T \Delta^{\mu\nu}\partial_{\nu} \left(\frac{\mu_R}{T}\right)\right\} = mh^{\mu\nu} \left( (E^n)_{\nu} + \frac{\partial_{\nu}T}{T}\right)\,,
\eeq
by virtue of $\mu_R = mc^2 +m \mu$.

We conclude this Subsubsection with an observation. While $C_{\mu}$ and $\mathcal{U}_{\mu}$ both have large $\mathcal{O}(c^2)$ pieces, there is a linear combination $C_{\mu} + m \mathcal{U}_{\mu}$ which has a regular $c\to\infty$ limit,
\beq
\lim_{c\to\infty} (C_{\mu} + m \mathcal{U}_{\mu}) = m\left\{ A_{\mu} +  u_{\mu} - \frac{1}{2}n_{\mu}u^2\right\} = m\tilde{A}_{\mu}\,,
\eeq
where $\tilde{A}_{\mu}$ is the Milne boost-invariant $U(1)$ connection in~\eqref{E:milneD}. Its field strength $\tilde{F}_{\mu\nu}$ yields the Milne boost-invariant electric and magnetic fields appearing in~\eqref{E:DuNR}.

\subsubsection{The non-relativistic Ward identities}
\label{S:NRwardFromLimit}

Intuitively, the relativistic conservation of charge and energy-momentum ought to imply that the non-relativistic number, energy, \&c are conserved. In flat space with no background electromagnetic field, this is almost immediate, but we would like to see how this works in general. That is, here we will show that the curved space relativistic Ward identities imply the curved space non-relativistic Ward identities upon taking $c\to\infty$.

Here and throughout the rest of this Section, we assume that spacetime derivatives commute with the large $c$ limit. This is reasonable in the backgrounds we study, where $(n_{\mu},h_{\mu\nu},A_{\mu})$ vary over spacetime in a way that does not scale with $c$. Under this assumption,
\beq
\label{E:numberWardFromR}
0 = m\lim_{c\to\infty} c\, \mathcal{D}_{\mu}j^{\mu} = \frac{m}{\sqrt{\gamma}}\partial_{\mu}\left(\sqrt{\gamma}\left\{ \lim_{c\to\infty} c j^{\mu}\right\}\right) = \left( \tilde{D}_{\mu} - \tilde{\mathcal{G}}_{\mu}\right)J^{\mu}\,,
\eeq
i.e. the large $c$ limit of the relativistic $U(1)$ Ward identity implies the non-relativistic Ward identity~\eqref{E:numberWard} for particle number. Similarly, the large $c$ limit of the relativistic stress tensor Ward identity gives
\begin{align}
\nonumber
\lim_{c\to\infty} c \left\{ \mathcal{D}_{\nu}t^{\mu\nu}-\mathcal{F}^{\mu}{}_{\nu}j^{\nu}\right\} & = \frac{1}{\sqrt{\gamma}}\partial_{\nu}\left( \sqrt{\gamma}\mathcal{T}^{\mu\nu}\right) +\Gamma^{\mu}{}_{\nu\rho}\mathcal{T}^{\nu\rho}+ \lim_{c\to\infty} c\left\{ \left( (\Gamma^R)^{\mu}{}_{\nu\rho}-\Gamma^{\mu}{}_{\nu\rho}\right) t^{\nu\rho} - \mathcal{F}^{\mu}{}_{\nu}j^{\nu}\right\}
\\
\label{E:stressWardFromLimit}
 & = ( D_{\nu}-\mathcal{G}_{\nu})\mathcal{T}^{\mu\nu} +\frac{1}{m} \lim_{c\to\infty} c\mathcal{F}^{\mu}{}_{\nu}\left\{ t^{\nu\rho}n_{\rho} - m j^{\nu}\right\}
\\
\nonumber
&= (D_{\nu}-\mathcal{G}_{\nu})\mathcal{T}^{\mu\nu} + (F^n)^{\mu}{}_{\nu}\mathcal{E}^{\nu}\,, \qquad \mathcal{G}_{\mu} = - F^n_{\mu\nu}v^{\nu}\,,
\end{align}
which is equal to the Milne boost-invariant version in~\eqref{E:NRward} upon adding zero in the right way~\cite{Jensen:2014ama}. In going from the second to the third line we have used
\begin{equation*}
\left(\Gamma^R\right)^{\mu}{}_{\nu\rho} - \Gamma^{\mu}{}_{\nu\rho} = \frac{1}{m}n_{(\nu}\mathcal{F}^{\mu}{}_{\rho)} + \mathcal{O}(c^{-2}) + (\text{torsion})\,.
\end{equation*}

To obtain the Ward identity for the energy current, we need recall our results for $\mathcal{D}_{\mu}\mathcal{U}_{\nu}$. We first simplify
\beq
\mathcal{U}_{\mu}\left( \mathcal{D}_{\nu}t^{\mu\nu} - \mathcal{F}^{\mu}{}_{\nu}j^{\nu}\right) + m c^2 \mathcal{D}_{\mu}j^{\mu} = \mathcal{D}_{\mu}\left( t^{\mu\nu}\mathcal{U}_{\nu} + m c^2 j^{\mu}\right) - t^{\mu\nu}\mathcal{D}_{\mu}\mathcal{U}_{\nu}+\mathcal{F}_{\mu\nu}j^{\mu}\mathcal{U}^{\nu}\,,
\eeq
and then using~\eqref{E:calEfromLimit},~\eqref{E:DUR},~\eqref{E:limitTheta}, and~\eqref{E:limitAccel} we find
\begin{align}
\begin{split}
\label{E:NRenergyFromLimit}
-\lim_{c\to\infty} c&\left\{ \mathcal{U}_{\mu}\left( \mathcal{D}_{\nu}t^{\mu\nu} - \mathcal{F}^{\mu}{}_{\nu}j^{\nu}\right) + m c^2 \mathcal{D}_{\mu}j^{\mu}\right\}
\\
& = -\frac{1}{\sqrt{\gamma}}\partial_{\mu}\left( \sqrt{\gamma}\lim_{c\to\infty}c\left\{ t^{\mu\nu}\mathcal{U}_{\nu} + mc^2j^{\mu}\right\}\right) + \lim_{c\to\infty} c\left\{ t^{\mu\nu}\mathcal{D}_{\mu}\mathcal{U}_{\nu}-\mathcal{F}_{\mu\nu}j^{\mu}\mathcal{U}^{\nu}\right\}
\\
& = \left( \tilde{D}_{\mu}-\tilde{\mathcal{G}}_{\mu}\right)\tilde{\mathcal{E}}^{\mu}+E^n_{\mu}\tilde{\mathcal{E}}^{\mu}  + \left( \sigma_{\mu\nu}+\frac{\vartheta}{d-1}\tilde{h}_{\mu\nu}\right)\mathcal{T}^{\mu\nu} - E_{\mu}J^{\mu}
\\
& = \left( \tilde{D}_{\mu} - 2\tilde{\mathcal{G}}_{\mu}\right) \tilde{\mathcal{E}}^{\mu} + \tilde{h}_{\rho(\mu}\tilde{D}_{\nu)}u^{\rho}\mathcal{T}^{\mu\nu}\,,
\end{split}
\end{align}
where the indices of $\sigma_{\mu\nu}$ have been lowered with $\tilde{h}_{\mu\nu}$ and we have used that $\mathcal{T}^{\mu\nu}n_{\nu} = J^{\mu}$. Of course, the last line is the energy Ward identity in~\eqref{E:TandE}.

\subsubsection{Ideal hydrodynamics}

Now let us see how ideal non-relativistic hydrodynamics emerges from the large $c$ limit of relativistic ideal hydrodynamics. As we justify more properly in Subsection~\ref{S:hydrostatic}, regularity of the large $c$ limit requires that the relativistic pressure $p$ satisfies
\beq
\label{E:pressureLimit}
\lim_{c\to\infty} c\,p(T,\mu_R) = P(T,\mu)\,,
\eeq
where $P$ will be the non-relativistic pressure. That is, the relativistic pressure goes as $\mathcal{O}(c^{-1})$. The relativistic energy density $\varepsilon_R$, entropy density $S$ and charge density $N$ are all determined from $p$ via
\beq
\varepsilon_R = - p + T S + \mu_R N\,, \qquad S = \left( \frac{\partial p}{\partial T}\right)_{\mu_R}\,, \qquad N = \left( \frac{\partial p}{\partial \mu_R}\right)_T\,.
\eeq
Using $\mu_R = mc^2 +m \mu$, these are related to the non-relativistic energy density $\varepsilon$, entropy density $s$, and charge density $\rho$ via
\begin{align}
\lim_{c\to\infty} c \, S &= s\,, &m \lim_{c\to\infty} c \, N &= \rho\,,
\\
\nonumber
 \lim_{c\to\infty} \frac{\varepsilon_R}{c} &=  \rho\,, & \lim_{c\to\infty} c(\varepsilon_R - mc^2 N) &= \varepsilon = - P + T s + \mu \rho\,.
\end{align}

The constitutive relations of ideal relativistic hydrodynamics are
\beq
\label{E:Rideal}
t^{\mu\nu} = \frac{\varepsilon_R }{c^2}\mathcal{U}^{\mu}\mathcal{U}^{\nu} + p \Delta^{\mu\nu}\,, \qquad j^{\mu} = N \mathcal{U}^{\mu}\,.
\eeq
Using the thermodynamic limits above together with~\eqref{E:calTfromLimit},~\eqref{E:calEfromLimit}, and~\eqref{E:DeltaLimit} we find
\begin{align}
\begin{split}
\mathcal{T}^{\mu\nu} = & \lim_{c\to\infty} c\, t^{\mu\nu} = \lim_{c\to\infty} \left( \frac{\varepsilon_R}{c}\mathcal{U}^{\mu}\mathcal{U}^{\nu} + c p \Delta^{\mu\nu}\right)
\\
 =&  \rho\, u^{\mu}u^{\nu} + P h^{\mu\nu}\,,
\\
\tilde{\mathcal{E}}^{\mu}  =&-\lim_{c\to\infty} c\left( t^{\mu\nu}\mathcal{U}_{\nu}+mc^2 j^{\mu}\right) =   \lim_{c\to\infty} c\left(  \varepsilon_R - m c^2N\right)\mathcal{U}^{\mu}
\\
= & \varepsilon u^{\mu}\,,
\end{split}
\end{align}
which are the constitutive relations of ideal non-relativistic hydrodynamics~\cite{LL6}, recast covariantly~\cite{Jensen:2014ama}.

\subsubsection{Beyond ideal hydrodynamics}
\label{S:beyondIdeal}

We conclude this Subsection with some schematic comments about how the large $c$ limit works beyond ideal hydrodynamics.

The most general relativistic constitutive relations can be parameterized as
\begin{align}
\begin{split}
t^{\mu\nu} & = \mathcal{E}^R \mathcal{U}^{\mu}\mathcal{U}^{\nu} + \mathcal{P}^R \Delta^{\mu\nu} + \mathcal{U}^{\mu}\mathcal{Q}^{\nu}+\mathcal{U}^{\nu}\mathcal{Q}^{\mu} + \mathscr{T}^{\mu\nu}\,,
\\
j^{\mu} & = \mathcal{N}^R\mathcal{U}^{\mu} + \mathscr{V}^{\mu}\,,
\end{split}
\end{align}
where $\mathcal{Q}^{\mu}, \mathscr{V}^{\mu},$ and $\mathscr{T}^{\mu\nu}$ are transverse to $ \mathcal{U}_{\mu}=0$ and $\mathscr{T}^{\mu\nu}$ is traceless. Regularity of the large $c$ limit requires that
\begin{align}
\nonumber
\mathcal{E}^R &= \frac{1}{c}\mathcal{N} + \frac{1}{c^3}\left( \mathcal{E} + \mathcal{N}_2\right)+ \hdots\,, & \mathcal{P}^R & = \frac{1}{c}\mathcal{P} + \hdots\,,
\\
\mathcal{N}^R & = \frac{1}{mc}\mathcal{N} + \frac{1}{mc^3}\mathcal{N}_2 + \hdots\,, & \mathcal{Q}^{\mu} & = \frac{1}{c}q^{\mu} + \frac{1}{c^3}\left( \eta^{\mu} + q_2^{\mu}\right) + \hdots\,,
\\
\nonumber
\mathscr{V}^{\mu} & = \frac{1}{mc}q^{\mu} + \frac{1}{mc^3}q_2^{\mu} + \hdots\,, &  \mathscr{T}^{\mu\nu} & = \frac{1}{c}\tau^{\mu\nu} + \hdots\,,
\end{align}
where the dots indicate terms that vanish faster than the last power of $c$, and the various scalars, vectors, and tensors here do not depend on $c$. The non-relativistic constitutive relations~\eqref{E:calTfromLimit} and~\eqref{E:calEfromLimit} are then
\beq
\tilde{\mathcal{E}}^{\mu} = \mathcal{E}u^{\mu}+\eta^{\mu}\,,
\qquad
\mathcal{T}^{\mu\nu} = \mathcal{N} u^{\mu}u^{\nu} + \mathcal{P} h^{\mu\nu} + u^{\mu}q^{\nu}+u^{\nu}q^{\mu} + \tau^{\mu\nu}\,.
\eeq

As we mentioned in~\eqref{E:limitTheta} and~\eqref{E:limitAccel}, the relativistic expansion, shear, and acceleration just become the non-relativistic expansion, shear, and ``energy electric field'' $E^n_{\mu}$ in the $c\to\infty $ limit. However, owing to the large $\mathcal{O}(c^2)$ terms in $C_{\mu}$ and $\mathcal{U}_{\mu}$, the electromagnetic fields and vorticity are typically $\mathcal{O}(c^2)$. One must then take care to ensure that the large $c$ limit is regular, that is that the non-relativistic energy current~\eqref{E:calEfromLimit}, \&c, are well-defined. Let us illustrate with two examples.

First, consider ordinary viscous hydrodynamics for a particular choice of the fluid variables known as Landau frame, with constitutive relations
\begin{align}
\begin{split}
\label{E:1stRhydro}
t^{\mu\nu} &= \frac{\varepsilon_R}{c^2}\mathcal{U}^{\mu}\mathcal{U}^{\nu} + \left( p - \zeta^R\vartheta^R\right)\Delta^{\mu\nu} - \eta^R (\sigma^R)^{\mu\nu}\,,
\\
j^{\mu} & = N \mathcal{U}^{\mu} + \sigma^R \left\{ \mathcal{F}^{\mu\nu}\mathcal{U}_{\nu} - T \Delta^{\mu\nu}\partial_{\nu}\left( \frac{\mu_R}{T}\right)\right\}\,.
\end{split}
\end{align}
Using that the relativistic expansion and shear become the non-relativistic expansion and shear, we see that we demand
\beq
\lim_{c\to\infty} c \zeta^R = \zeta\,, \qquad \lim_{c\to\infty} c \eta^R = \eta\,,
\eeq
and then~\eqref{E:calTfromLimit} gives
\beq
\mathcal{T}^{\mu\nu} =  \rho \,\mathcal{U}^{\mu}\mathcal{U}^{\nu} + (P - \zeta \vartheta)h^{\mu\nu} - \eta \sigma^{\mu\nu}\,.
\eeq
The tensor structure multiplying the relativistic conductivity $\sigma^R$ is $\mathcal{O}(c^2)$,~\eqref{E:limitV}. In order to have a well-defined energy current,~\eqref{E:aCaseOfTheObvious} implies that $\sigma^R$ is at least $\mathcal{O}(c^{-5})$, and then we find that the Milne boost-invariant energy current~\eqref{E:calEfromLimit} is
\beq
\tilde{\mathcal{E}}^{\mu} = \varepsilon \mathcal{U}^{\mu} -m^2 \sigma \left( (E^n)^{\mu} + \frac{\partial^{\mu}T}{T}\right)\,, \qquad \lim_{c\to\infty}c^5\sigma^R=\sigma\,,
\eeq
from which we find that the thermal conductivity, the transport coefficient multiplying $-\partial^{\mu}T$, is $\kappa = m^2\sigma/T$. This is just covariant first-order non-relativistic hydrodynamics in Eckhart frame~\cite{Jensen:2014ama}. One can work in an arbitrary relativistic fluid frame provided that the large $c$ limit is regular, in which case one will get first-order non-relativistic hydrodynamics in an arbitrary fluid frame.

Our second example involves parity-violating first-order hydrodynamics in $2+1$ dimensions~\cite{Jensen:2011xb}. A complete treatment of the non-relativistic limit of this hydrodynamics is beyond the scope of this work, but here we simply want to give a taste of how the analysis should work by focusing on the potential contributions of pseudoscalars. At one-derivative order, there are two pseudoscalars,
\beq
B = - \frac{1}{2}\varepsilon^{\mu\nu\rho} \mathcal{U}_{\mu}\mathcal{F}_{\nu\rho}\,, \qquad \Omega = - \varepsilon^{\mu\nu\rho}\mathcal{U}_{\mu}\partial_{\nu}\mathcal{U}_{\rho}\,.
\eeq
At large $c$, both are $\mathcal{O}(c^3)$,
\beq
\lim_{c\to\infty} \frac{B}{c^3} =  m \varepsilon^{\mu\nu\rho}n_{\mu}\partial_{\nu}n_{\rho}\,, \qquad \lim_{c\to\infty}\frac{\Omega}{c^3} = - \varepsilon^{\mu\nu\rho}n_{\mu}\partial_{\nu}n_{\rho}\,.
\eeq
So one simple option is that the response coefficients multiplying $B$ and $\Omega$ are $\mathcal{O}(c^{-4})$, so that $ct^{\mu\nu}$ and $cj^{\mu}$ are regular as $c\to\infty$. Then there is only one independent non-relativistic pseudoscalar one finds in the large $c$ limit, namely
\begin{equation*}
\mathcal{B}^n = \varepsilon^{\mu\nu\rho}n_{\mu}\partial_{\nu}n_{\rho}\,.
\end{equation*}
The other simple possibility is that $B$ and $\Omega$ appear together through the $\mathcal{O}(c)$ combination $B+m \Omega$, in which case the corresponding response coefficients ought to be no larger than $\mathcal{O}(c^{-2})$. The most general option is that $B$ and $\Omega$ appear in $t^{\mu\nu}$ and $j^{\mu}$ through
\begin{equation*}
\frac{1}{mc^2}\left\{f_1(T,\mu) + \frac{f_2(T,\mu)}{c^2} + \hdots \right\}B + \frac{1}{c^2}\left\{ f_1(T,\mu) + \frac{f_3(T,\mu)}{c^2}+ \hdots\right\} \Omega\,,
\end{equation*}
where the dots vanish vanish faster than $\mathcal{O}(c^{-2})$. The $c\to\infty$ limit then gives
\begin{equation*}
f_1 \mathcal{B} + (f_2 - f_3)\mathcal{B}^n\,, \qquad \mathcal{B} = \frac{1}{2}\varepsilon^{\mu\nu\rho}n_{\mu}\tilde{F}_{\nu\rho}\,.
\end{equation*}
In non-relativistic first-order hydrodynamics, $\mathcal{B}$ and $\mathcal{B}^n$ are the two one-derivative pseudoscalars. So we see that an appropriate large $c$ limit of $B$ and $\Omega$ yields an arbitrary linear combination of $\mathcal{B}$ and $\mathcal{B}^n$.

In any case, we hope that the moral is clear: one must carefully scale the powers of $c$ appearing in transport coefficients in order to ensure that there is a large $c$ limit.

%%%%%%%%%%%%%%%%%%%%%%%%%%%%%%%%%%%%%%%%%
\subsection{Entropy}
\label{S:entropy}
%%%%%%%%%%%%%%%%%%%%%%%%%%%%%%%%%%%%%%%%%

In the beginning of this Section, we mentioned that the constitutive relations of hydrodynamics are not arbitrary: they must be consistent with a local version of the second Law. That is, for fluid flows which solve the hydrodynamic equations, there is an entropy current $S^{\mu}$ which in the flat-space equilibrium is $S^{\mu} = S \mathcal{U}^{\mu}$ and which satisfies
\beq
\mathcal{D}_{\mu}S^{\mu} \geq 0\,.
\eeq
It is of course equivalent to demand (see e.g.~\cite{Loganayagam:2011mu})
\beq
\label{E:entropy}
T\mathcal{D}_{\mu}S^{\mu} + \mu_R \mathcal{D}_{\mu}j^{\mu} +\mathcal{U}_{\mu}\left( \mathcal{D}_{\nu}t^{\mu\nu}-\mathcal{F}^{\mu}{}_{\nu}j^{\nu}\right)\geq 0\,,
\eeq
as we have just added the Ward identities which vanish ``on-shell.'' It is implicit in writing~\eqref{E:entropy} that one solves~\eqref{E:entropy} without using the hydrodynamic equations. For various reasons,~\eqref{E:entropy} is a more useful version of the entropy criterion.

There is of course a non-relativistic version of this story~\cite{Jensen:2014ama}. We demand the existence of an entropy current $s^{\mu}$ satisfying
\beq
\label{E:NRentropy}
T \left( \tilde{D}_{\mu}-\tilde{\mathcal{G}}_{\mu}\right)s^{\mu} + \mu \left( \tilde{D}_{\mu}-\tilde{\mathcal{G}}_{\mu}\right)J^{\mu} - \left( \tilde{D}_{\mu}-2\tilde{\mathcal{G}}_{\mu}\right)\tilde{\mathcal{E}}^{\mu} - \tilde{h}_{\rho(\mu}\tilde{D}_{\nu)}u^{\rho}\mathcal{T}^{\mu\nu} \geq 0\,.
\eeq
As above, we have just added a linear combination of the particle number and energy Ward identities to the divergence of the entropy current.

Earlier, we showed that the relativistic Ward identities imply the non-relativistic Ward identities in the $c\to\infty$ limit. One might then expect that if the relativistic entropy criterion~\eqref{E:entropy} is satisfied, then the non-relativistic one will be too. Indeed, using~\eqref{E:NRenergyFromLimit} and $\mu_R = m c^2 + \mu$, we see that this is the case with
\beq
\lim_{c\to\infty}c\,S^{\mu} = s^{\mu}\,.
\eeq

%%%%%%%%%%%%%%%%%%%%%%%%%%%%%%%%%%%%%%%%%
\subsection{The hydrostatic partition function}
\label{S:hydrostatic}
%%%%%%%%%%%%%%%%%%%%%%%%%%%%%%%%%%%%%%%%%

The local second Law, as described above, imposes constraints on the transport coefficients which appear in the constitutive relations of hydrodynamics. The constraints fall into two types, equality-type and inequality-type. A simple example of an equality-type constraint is already visible in the constitutive relation for the current in~\eqref{E:1stRhydro}:  the local electric field $\mathcal{F}^{\mu\nu}\mathcal{U}_{\nu}$ only appears through the combination $\mathcal{F}^{\mu\nu}\mathcal{U}_{\nu} - T \Delta^{\mu\nu}\partial_{\nu}\left( \frac{\mu_R}{T}\right)$, which amounts to the Einstein relation between electric and thermal conductivities. An inequality-type relation is just that the electric conductivity $\sigma^R$ must be non-negative.

Recently, it has been understood how the equality-type relations are a consequence of symmetries. As explained in~\cite{Jensen:2012jh,Banerjee:2012iz} for relativistic thermal field theory (see also~\cite{Jensen:2012jy}) and~\cite{Jensen:2014ama} for Galilean systems, the thermal partition function of a theory on a hydrostatic spacetime background simplifies dramatically compared to the full partition function. By hydrostatic, we mean that the background metric and gauge fields are time-independent, but vary slowly over long distances. The hydrostatic partition function is just the functional integral on an appropriate Euclidean version of the time-independent spacetime, in which case the thermal circle is much smaller than the variations on the spatial slice. One can imagine reducing on the thermal circle. When the microscopic theory has finite correlation length, which is almost always the case at $T>0$, the effective description on the spatial slice is a gapped field theory. It immediately follows that the hydrostatic partition function can be written locally on the spatial slice in a gradient expansion of the spacetime background. Each term in the gradient expansion is an integral of a gauge-invariant scalar.

One can then classify the terms that appear in the thermal partition function to any fixed order in gradients. Varying the partition function, one finds all Euclidean zero-frequency correlators. Matching these correlators to hydrodynamics, one finds in every example considered thus far that the two can be matched only if the equality-type relations are satisfied. See also~\cite{Bhattacharyya:2013lha} for strong evidence that this is always the case.

It will probably not shock the reader that we can also obtain the NR hydrostatic partition function from a large $c$ limit. Let us sketch how it works. Rather than working with a gradient expansion on the spatial slice, we use the covariant analysis of~\cite{Jensen:2012jh,Jensen:2014ama}, in which the partition function is written in a gradient expansion on the Euclidean spacetime, and the scalars appearing therein are built from the background and the symmetry data. The covariant version of the statement that the background is time-independent is that there is a timelike vector field $K^{\mu}$ and gauge transformation $\Lambda_K$ which generate a symmetry of the background. Denoting them as $K=(K^{\mu},\Lambda_K)$, we mean that
\beq
\label{E:RdeltaK}
\delta_K g_{\mu\nu} = \pounds_K g_{\mu\nu} = 0\,, \qquad \delta_K C_{\mu} = \pounds_K C_{\mu} +m \partial_{\mu}\Lambda_K = 0\,.
\eeq
Picking coordinates and a gauge so that $K^{\mu}=\delta^{\mu}_t$ and $\Lambda_K=0$,~\eqref{E:RdeltaK} just means that $g_{\mu\nu}$ and $A_{\mu}$ are independent of $t$. Let us normalize $K^{\mu}$ and $\Lambda_K$ so that they are $\mathcal{O}(1)$ in the large $c$ limit. The Euclidean spacetime is just built from Wick-rotating the affine parameter along the integral curves of $K^{\mu}$, and compactifying with imaginary periodicity $\beta$. From $K^{\mu}$ and $\Lambda_K$ we can construct a local temperature $T$, fluid velocity $\mathcal{U}^{\mu}$, and chemical potential $\mu_R$ via
\beq
\label{E:RsymmetryData}
T= \frac{c}{\beta \sqrt{-K^2}}\,, \qquad \mathcal{U}^{\mu} = \frac{c K^{\mu}}{\sqrt{-K^2}}\,, \qquad \frac{\beta\mu_R}{T}=K^{\mu}C_{\mu} + m\Lambda_K\,.
\eeq
The relativistic hydrostatic partition function can then be written down in a gradient expansion of gauge-invariant scalars built from the background fields, $(T,\mathcal{U}^{\mu},\mu_R)$, and the covariant derivative.

Taking the large $c$ limit of $(T,\mu_R,\mathcal{U}^{\mu})$, we find
\beq
\lim_{c\to\infty}T = \frac{1}{\beta \, n_{\mu}K^{\mu}}=T_{NR}\,, \qquad \lim_{c\to\infty} \mathcal{U}^{\mu} = \frac{K^{\mu}}{n_{\nu}K^{\nu}}=u^{\mu}\,,
\eeq
where we abuse notation and refer to the large $c$ limits of $K^{\mu}$ and $\Lambda_K$ as $K^{\mu}$ and $\Lambda_K$. These are exactly the local temperature and fluid velocity identified in~\cite{Jensen:2014ama} for a Galilean field theory coupled to a Newton-Cartan background. We also have
\beq
\mu_R = mc^2 + m\mu\,, \qquad \lim_{c\to\infty} \mu = \frac{K^{\mu}\tilde{A}_{\mu} + \Lambda_K}{n_{\nu}K^{\nu}}=\mu_{NR}\,,
\eeq
where $\mu_{NR}$ is exactly the local chemical potential for a Galilean theory. Henceforth, we drop the `$NR$'s and simple refer to the NR temperature and chemical potential as $T$ and $\mu$.

The hydrostatic generating functional
\beq
W^R_{hydrostat} = - i \ln \mathcal{Z}^R_{hydrostat}\,,
\eeq
can be ``un-Wick-rotated'' in such a way that it can be written as a functional of the original background,
\beq
W^R_{hydrostat} = W^R_0 + W^R_1 + \hdots\,, \qquad W^R_n= \int d^dx \sqrt{-g} \mathcal{L}_n\,,
\eeq
where $\mathcal{L}_n$ is a gauge-invariant $\mathcal{O}(\partial^n)$ scalar, and the integral is understood to be performed over the Euclidean spacetime.\footnote{The reason for the gymnastics with analytic continuation is simply that writing $W^R_{hydrostat}$ this way makes it easy to compute the real-time hydrostatic response. One simply varies $W^R_{hydrostat}$ with respect to the background fields, without having to introduce any factors of $i$.} The zeroth order term is just
\beq
W^R_0 = \int d^dx \sqrt{-g} p(T,\mu_R)\,,
\eeq
and its variations give the stress tensor and current of ideal hydrodynamics~\eqref{E:Rideal}. Regularity of the large $c$ limit means that $\lim_{c\to\infty} W^R_0$ must exist, in which case $p$ must scale as $\mathcal{O}(c^{-1})$. That is, the limit only exists if~\eqref{E:pressureLimit} holds, which derives that condition, in which case
\beq
\lim_{c\to\infty} W_0^R = W_0=\int d^dx \sqrt{\gamma} P(T,\mu)\,,
\eeq
which is indeed the zero derivative term in the NR hydrostatic generating functional~\cite{Jensen:2014ama}.

It turns out that $W^R_1$ vanishes in parity-preserving theories, which corresponds to the fact that there is no parity-preserving, dissipationless transport in first-order hydrodynamics. The same is true in NR theories~\cite{Jensen:2014ama}: $W_1$ vanishes in parity-preserving theories, which matches first-order hydrodynamics. When parity is broken, $W_1$ may be nonzero. In two spatial dimensions, there are two pseudoscalars one can form from the spacetime background and symmetry data, $B$ and $\Omega$, and so in that case
\beq
W_1^R = \int d^3x \sqrt{-g} \left\{ \tilde{f}_1(T,\mu_R) B + \tilde{f}_2(T,\mu_R)\Omega\right\}\,.
\eeq
As we discussed in Subsection~\ref{S:beyondIdeal}, $B$ and $\Omega$ are $\mathcal{O}(c^3)$ at large $c$, but the combination $B+m\Omega$ is $\mathcal{O}(c)$. Consequently, the large $c$ limit implies that
\beq
\tilde{f}_1 = \frac{1}{mc^2}\left\{ f_1(T,\mu) + \frac{1}{c^2}f_2(T,\mu) + \hdots\right\}\,, \qquad \tilde{f}_2 = \frac{1}{c^2}\left\{  f_1(T,\mu) + \frac{1}{c^2}f_3(T,\mu) + \hdots\right\}\,,
\eeq
where the dots vanish faster than $\mathcal{O}(c^{-2})$. Then
\beq
\lim_{c\to\infty} W_1^R = W_1=\int d^3x \sqrt{\gamma}\left\{ f_1(T,\mu)\mathcal{B} + \left(f_2(T,\mu)-f_3(T,\mu)\right) \mathcal{B}^n\right\}\,,
\eeq
where $\mathcal{B} =\frac{1}{2} \varepsilon^{\mu\nu\rho}n_{\mu}\tilde{F}_{\nu\rho}$ and $\mathcal{B}^n = \varepsilon^{\mu\nu\rho}n_{\mu}\partial_{\nu}n_{\rho}$ are the NR boost-invariant magnetic field and ``energy magnetic field.'' These are two allowed one-derivative terms that can appear in the NR $W_1$~\cite{Jensen:2014ama}, and so one can obtain the most general one-derivative term in the NR $W$ from a limit of the most general relativistic $W_1^R$. We do not know if it is always the case that a general NR hydrostatic $W$ can be obtained from a large $c$ limit, or if the limit places constraints on other transport allowed by NR symmetry. 

%%%%%%%%%%%%%%%%%%%%%%%%%%%%%%%%%%%%%%%%%
\section{Conclusions}
\label{S:conclude}
%%%%%%%%%%%%%%%%%%%%%%%%%%%%%%%%%%%%%%%%%

In this work we have demonstrated how to obtain the full NC data describing a NR theory on a generic curved background from a NR limit of a relativistic parent. Most importantly, the Milne redundancy of the NC data naturally arises from this limit. We have also confirmed that our construction nicely reproduces the known modifications of the Milne boosts in the presence of magnetic moment terms.

We have focused on two simple examples of a relativistic parent, perturbative scalars and hydrodynamics. It would be very interesting to implement our procedure for different Lagrangians. In particular, in order to describe the low energy effective action of gapped systems in 2+1 dimensions it would be imperative to take an NR limit of the gauge and gravitational Chern-Simons terms see~\cite{Andreev:2013qsa} for such a computation when $n_{\mu}$ is constant). Another possible generalization of interest is to study the NR limit of Dirac fermions in curved space. While technically somewhat cumbersome, these exercises should be straight forward given the tools developed here.

One further potential application of our results is holographic. In \cite{Janiszewski:2012nb} it was argued that if a relativistic parent has a dual description in terms of standard Einstein gravity, the same large $c$ limit outlined here can also be implemented in the gravity dual. One turns on a background gauge field $C_{\mu}$ in an asymptotically anti-de Sitter (AdS) background whose boundary value is given by the same $C_{t}=mc^2$ used in this work for the special case that $n_{\mu}$ points only in the time direction. As argued in \cite{Janiszewski:2012nb}, a constant $C_t$ in the bulk cannot be gauged away at non-zero charge density. If we now take the $c \rightarrow \infty$ limit in the bulk, most degrees of freedom decouple. \cite{Janiszewski:2012nb} used symmetries to argue that the resulting theory is a variant of Horava gravity \cite{Horava:2009uw}. If we insist to only mod out by diffeomorphisms in the bulk that leave the leading $\mathcal{O}(c^2)$ term in $C_t$ invariant, we are left with time-dependent spatial diffeomorphisms as well as time reparametrizations, the defining symmetries of Horava gravity. Standard Horava gravity can be rewritten as Einstein gravity coupled to a scalar field, the khronon \cite{Germani:2009yt,Blas:2009yd}, whose time gradient picks a preferred time direction. Instead of this scalar, the preferred time direction in holographic construction described in \cite{Janiszewski:2012nb} is fixed by $C_t$, so that theory was referred to as Horava gravity with a ``vector khronon.'' An explicit string theory embedding of this scenario is just a null reduction of pure AdS~\cite{Janiszewski:2012nf}. Of course we can restore the full relativistic diffeomorphism invariance if we let the leading term in $C_{\mu}$, which on the boundary gives $n_{\mu}$, transform non-trivially. The field theory analysis performed in this work shows that in this case the boundary theory is best described in terms of NC data redundant under Milne boosts, and so the same is true about the bulk theory dual to this limiting procedure. However, it is not clear to us how or even if the Milne invariance is realized in Horava gravities apart from this limit.

Other than the physical interpretation of the Milne boosts, maybe the most important insight gained from our construction is an answer to the question of when a NR theory should have a good relativistic parent. At least, we found an answer when the NR theory is just realized as the large $c$ limit of a perturbative relativistic theory. We saw that demanding that a NR theory comes from a limit from a relativistic parent restricted the NR couplings beyond what is imposed by the NR symmetries. For example, in two spatial dimensions, there are two NR magnetic moments that are consistent with the NR symmetries. However, only one of them is realized from the large $c$ limit. We traced this difference to the fact that in the NR theory one does not demand invariance under gauge transformations that shift the leading $\mathcal{O}(mc^2)$ term in the relativistic background gauge field.

As all NR systems in our world do in fact follow as NR limits of an underlying relativistic theory, one may wonder if they should obey the strong requirement of following from a reduction procedure as outlined here. However, we note that  we only performed our construction at the free field level as well as in the hydrodynamic limit. We can only argue for this more stringent constraint on the daughter theory in the case that both the parent and the daughter are free theories (or small perturbations thereof) or within the hydrodynamic regime. If our results were to continue to hold in the case of generic interacting theories, this would put additional strong constraints on, say, the low energy effective action of Hall systems that can appear in nature. Not every Galilean boost invariant low energy effective action that can be written on a piece of paper would actually be realizable as a physical theory. 

\acknowledgments

We are pleased to thank J.~Fuini, S.~Janiszewski, M.~Kaminski, M.~Pospelov, C.~Uhlemann, and D.~Son for useful discussions. The work of KJ was supported in part by National Science Foundation under grant PHY-0969739. The work of AK was supported, in part, by the US Department of Energy under grant number DE-SC0011637.

\bibliographystyle{JHEP}
\bibliography{NRlimitwithg}

\end{document}